\documentclass[pdflatex,sn-nature]{sn-jnl}% Style for submissions to Nature Portfolio journals
\usepackage{graphicx}%
\graphicspath{{articlefigure/}}
\usepackage{multirow}%
\usepackage{amsmath,amssymb,amsfonts}%
\usepackage{amsthm}%
\usepackage{mathrsfs}%
\usepackage[title]{appendix}%
\usepackage{xcolor}%
\usepackage{textcomp}%
\usepackage{manyfoot}%
\usepackage{booktabs}%
\usepackage{url}
\usepackage{algorithm}%
\usepackage{algorithmicx}%
\usepackage{algpseudocode}%
\usepackage{listings}%
\usepackage{setspace}
\usepackage[version=4]{mhchem}
\usepackage{xr}
\theoremstyle{thmstyleone}%
\theoremstyle{thmstyletwo}%

\theoremstyle{thmstylethree}%

\begin{document}

\title[Article Title]{Water structuring at stacked graphene interfaces unveiled by machine-learning molecular dynamics}


\author[1,2]{\fnm{Dianwei} \sur{Hou}}\email{davyhou@korea.ac.kr}
\author[1,2]{\fnm{Yevhen} \sur{Horbatenko}}\email{horbatenko@korea.ac.kr}
\author*[1,2]{\fnm{Stefan} \sur{Ringe}}\email{sringe@korea.ac.kr}
\author*[1,2]{\fnm{Minhaeng} \sur{Cho}}\email{mcho@korea.ac.kr}

\affil[1]{\orgdiv{Center for Molecular Spectroscopy and Dynamics}, \orgname{Institute for Basic Science (IBS)}, \city{Seoul}, \postcode{02841}, \country{Republic of Korea}}

\affil[2]{\orgdiv{Department of Chemistry}, \orgname{Korea University}, \city{Seoul}, \postcode{02841}, \country{Republic of Korea}}


\abstract{
The wettability of monolayer and multilayer graphene remains a topic of longstanding debate. Here, we combined first-principles molecular dynamics simulations accelerated with the atomic cluster expansion machine learning interatomic potential to investigate how substrate, graphene layer number, and intercalated water molecules influence graphene’s wettability. Simulated vibrational sum-frequency generation (vSFG) spectra revealed that the experimentally observed hydrophilic behavior of monolayer graphene on hydrophilic substrates arose not from wetting transparency, but from signal cancellation induced by intercalated water. Energetic analyses further showed that intercalated water molecules were thermodynamically favorable for monolayer graphene on hydrophilic substrates, but not for multilayer systems, leading to changes in the vSFG response in line with experimental observations. These results offer a mechanistic understanding of graphene-water interactions and have broad implications for the design of graphene-based interfaces and devices.
}

\maketitle

\section{Introduction}

Since the successful mechanical exfoliation of graphene in 2004,\cite{RN592} two-dimensional (2D) materials have garnered widespread scientific interest owing to their unique physical and chemical properties, which are fundamentally distinct from those of their bulk counterparts.\cite{RN625, RN624, RN596} Graphene, a single layer of carbon atoms arranged in a hexagonal lattice with $\mathrm{sp^2}$ hybridization, is one of the most representative 2D materials.\cite{RN648} It possesses a combination of remarkable properties, including high electrical and thermal conductivity, excellent optical transparency, superior chemical stability, outstanding mechanical strength, and complete impermeability to atoms and molecules.\cite{RN629, RN630, RN627, RN628, XU201873, RN631} These remarkable properties have enabled a wide range of applications, such as desalination of water, transparent conductive electrodes, electrical energy storage, and electrocatalysis.\cite{RN646, RN643, RN647, RN644, RN645} Notably, many of these applications involve the interaction of graphene with water or aqueous electrolytes.\cite{RN641, RN642, RN53} This interaction enforces a distinct molecular structure of water, expressed through orientation of water molecules and the hydrogen-bond network, both of which are crucial for the optimal performance of devices like proton exchange membrane fuel cells\cite{B808149M, moilanen2008water, xu2019transparent} and nanofluidic devices\cite{sachar2021hydrogen, joly2016strong}. 

One of the most widely used experimental techniques for evaluating graphene-water interactions is the measurement of the water contact angle (WCA).\cite{RN609, RN607, RN608} However, WCA values are influenced by a variety of factors, including the presence of an underlying substrate,\cite{RN586, RN499, RN496} the defect density on both the graphene and the substrate,\cite{RN616, RN615, RN612, RN613} the number of graphene layers,\cite{RN91, RN496} and surface contaminants\cite{doi:10.1021/acs.jpcc.5b10492, RN599, RN614}. These factors can result in a wide range of reported WCA values, ranging from $30^\circ$ to $127^\circ$.\cite{RN619, RN618, RN600} Given that materials with WCA below $90^\circ$ are considered hydrophilic and those above $90^\circ$ hydrophobic, there remains ongoing debate as to whether graphene is intrinsically hydrophilic or hydrophobic.\cite{BELYAEVA2020100482} Although WCA measurement is a useful tool for characterizing surface wettability, it is a macroscopic observable and does not directly reveal molecular-level details of the interfacial water structure. Therefore, a comprehensive understanding of the wettability of graphene requires molecular-scale insight into the graphene-water interactions that govern its observed WCA behavior.

Vibrational sum-frequency generation (vSFG) spectroscopy\cite{RN621, RN620, RN26, RN15, RN622} is a powerful technique for probing the molecular structure of water at interfaces.\cite{RN606, RN103, RN605, RN604, RN603, RN602, RN107} As a second-order non-linear optical process, vSFG requires breaking the inversion symmetry to generate a finite signal. This requirement makes the technique inherently surface-specific, as bulk water is centrosymmetric and thus does not contribute to the optical response. Consequently, the vibrational signals detected by vSFG arise exclusively from interfacial water molecules. Experimental and simulation results from Wang et al.\cite{RN493}, along with independent experimental measurements by Xu et al.\cite{RN669}, both demonstrated that pristine graphene floating on water exhibited hydrophobic behavior. In contrast, when supported on a calcium fluoride (\ce{CaF2}) substrate, the measured vSFG spectra indicated a hydrophilic character of the graphene layer.\cite{RN97, RN551, RN91, RN525} Given the hydrophilicity of the substrate and the atomic thickness of graphene, some researchers proposed that the wettability of the underlying substrate significantly influences the overall interfacial properties of graphene and water. This phenomenon was referred to as wetting transparency and was proposed to explain the observed wettability of the graphene surface.\cite{RN499, RN496} However, the extent to which graphene is transparent to wetting remains a subject of ongoing debate. Graphene has been reported to be fully wetting-transparent,\cite{RN91, RN496, RN503, RN504} partially wetting-transparent,\cite{RN598, RN500, RN599, RN611} or fully wetting-opaque\cite{RN589, RN501, RN601, RN600, D0NR08843A}. 

In addition to substrate effects, intercalated water molecules may also influence the shape of vSFG spectra. Ohto et al.\cite{RN52} simulated the vSFG spectra of pristine graphene and found that when all water molecules were located on one side of the graphene sheet, the spectrum closely resembled that of the water-air interface, consistent with the hydrophobic nature of graphene. However, the presence of even a small number of water molecules on the opposite side of the graphene layer significantly reduced the peak amplitude of the dangling O-H bond, indicating a diminished hydrophobic response. In experimental work, Temmen et al.\cite{RN515, RN74} used non-contact atomic force microscopy and Kelvin probe force microscopy to characterize the thickness of graphene sheets and the intercalated water layer. They found that even after annealing to 750 K, it was nearly impossible to completely remove the water molecules trapped between graphene and the substrate \ce{CaF2}. Li et al.\cite{RN528} investigated the behavior of intercalated water by calculating the migration energy barrier of a single water molecule confined between monolayer graphene and substrate \ce{CaF2}, using the climbing image nudged elastic band method. They reported a migration barrier of approximately 0.19 eV. They also calculated the diffusion coefficient of water at 300 K, finding it to be 8.35 × \ensuremath{10^{-6}} \text{cm}$^2$/s. These results, a low migration barrier, a high diffusion rate, and the persistence of water even after high-temperature annealing, collectively suggest the presence of intercalated water between graphene and the \ce{CaF2} substrate.

Beyond substrate effects and intercalated water, the number of graphene layers can also influence the experimentally measured vSFG spectra. Kim et al.\cite{RN91} systematically investigated the WCA and vSFG spectra of multilayer graphene. Their vSFG measurements revealed that when the number of graphene layers exceeded four, the vSFG spectra exhibited pronounced changes. These changes were attributed to the transition of the graphene film from wetting transparent to translucent and finally to opaque as the number of layers increases from one to six. However, the fundamental mechanisms underlying these dramatic changes in the vSFG response with increasing graphene thickness were not fully understood.

Although experimental vSFG provides valuable information about interfacial water structure, it remains challenging to control atomically the interface to disentangle the various factors, such as graphene-water interactions, graphene layer effects, substrate influence, and water penetration. To achieve a comprehensive microscopic understanding, atomic-scale simulations are indispensable. Ab initio molecular dynamics (AIMD) simulations can provide such insights with quantum-chemical accuracy that allows for a reliable description of the water orientation and hydrogen-bond network at the graphene surface. Unfortunately, AIMD simulations are generally limited by high computational cost and insufficient sampling. Accurate simulation of vSFG spectra often requires nanosecond time-scale trajectories to ensure statistical convergence.\cite{RN1, RN157} To solve this issue, machine learning interatomic potentials (MLIP) have recently become impressively successful in interpolating the quantum-chemical potential energy surface, reaching AIMD-level accuracy while dramatically improving computational efficiency.

Accurate simulation of the water structure at the graphene-water interface is a complex task. The orientation-dependent adsorption energy of single water on graphene varies by less than 0.043 eV (1 kcal/mol), necessitating simulations that exceed the threshold of “chemical accuracy”.\cite{RN43, RN80, RN35, RN623, RN591} Moreover, the graphene-water interactions are modulated by several factors, including the substrate,\cite{RN43, RN496} the number of graphene layers,\cite{RN91, munz2015thickness} the defect of graphene,\cite{RN613} the thickness of the water slab,\cite{RN35, RN80}, external electrical field,\cite{RN591, RN623, D3FD00107E, RN669} and the presence of intercalated water molecules between the graphene and the substrate.\cite{RN62, RN60, RN662} Capturing such a complex, multi-element, and many-atom system with sub-chemical accuracy pushes the capabilities of MLIP beyond what has been explored in the existing literature. In addition, previous theoretical studies of vSFG spectra at graphene-water interfaces have primarily focused on simplified systems, such as monolayer graphene with a few water molecules,\cite{RN52, RN493, wang2025spectralsimilaritymasksstructural, du2025machinelearningacceleratedcomputational} or systems with the \ce{CaF2} substrate and a monolayer graphene containing a small number of intercalated water molecules.\cite{RN662} 

In this work, we develop chemically accurate MLIPs based on the Atomic Cluster Expansion (ACE) method\cite{RN4, RN22}, trained in extensive density functional theory (DFT) calculations, and applied it to long-term large-scale MD simulations. These simulations covered graphene systems with varying numbers of layers, both with and without the substrate \ce{CaF2}, and with different numbers of intercalated water molecules. The resulting MD trajectories were used to investigate the orientation of water molecules at the graphene-water interface and to simulate the corresponding vSFG spectra. From these results, we clarified the role of the substrate, the number of graphene layers, and intercalated water in controlling the interfacial water structure, as well as the observed vSFG spectra. The simulated spectra exhibit good agreement with the experimental observations and offer new insights into their interpretation. Notably, our simulations suggest a potential misinterpretation of previously reported vSFG spectra of monolayer graphene on hydrophobic substrates. They further support the perspective that graphene behaves as hydrophobic and non-wetting transparent, with this behavior appearing to be largely insensitive to the number of stacked layers. Furthermore, our simulations highlight the impressive performance of current MLIPs when tested in challenging system setups.

\begin{figure}[ht]
\centering
\includegraphics[width=0.8\textwidth]{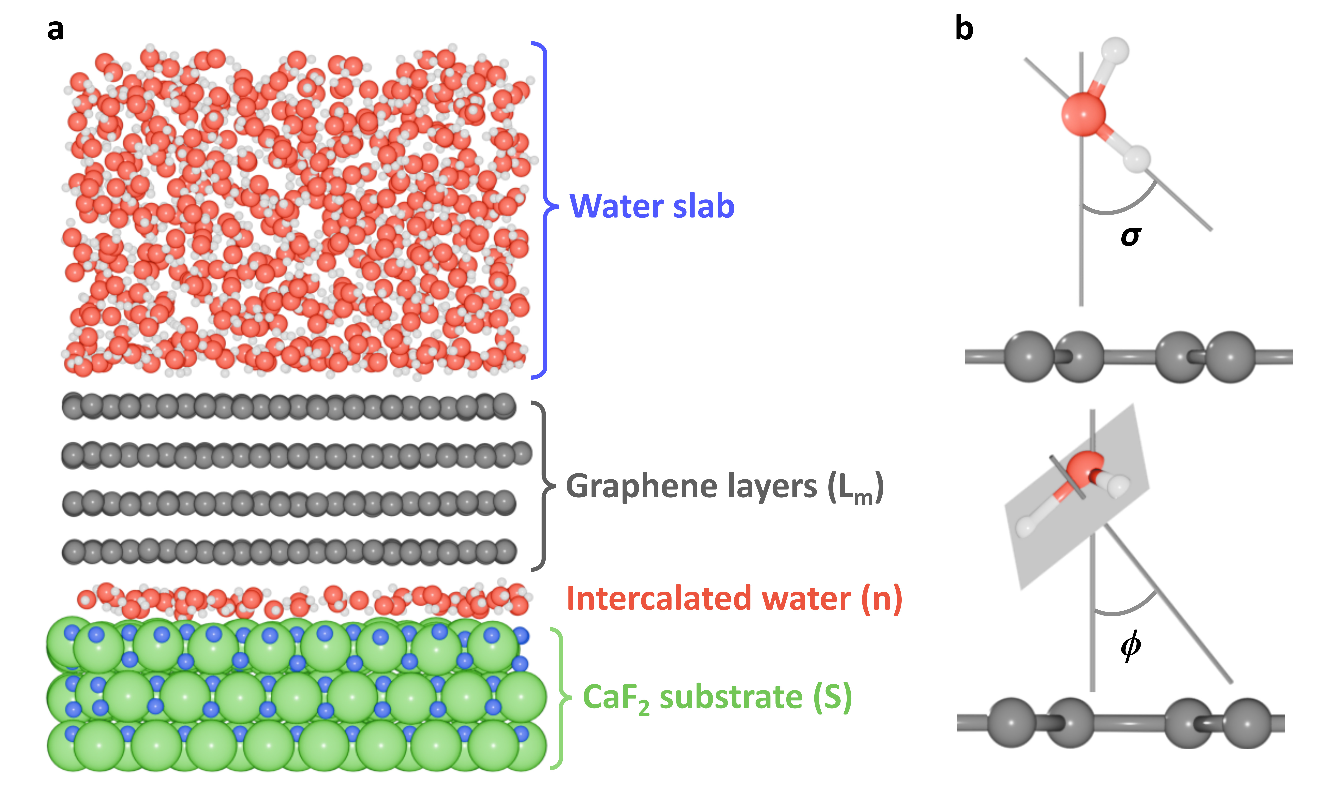}
\caption{\textbf{a}, Atomistic model of the $S_{48}L_4$ system, comprising a \ce{CaF2} substrate with 48 intercalated water molecules, four graphene layers, and an overlying water slab. \textbf{b}, Angular distribution of water molecules, defined by two orientation angles: $\sigma$, representing the angle between the lower-coordinated O-H bond and the Z-axis; and $\phi$, representing the angle between the normal vector of the water molecular plane and the Z-axis.
%\textbf{b}, A schematic of the workflow that combines the training of a machine learning potential using the atomic cluster expansion (ACE) framework with iterative refinement driven by an active learning strategy. 
}
\label{fig1}
\end{figure}

\section{Results}

\subsection{Atomistic models} \label{subsec2.1}

To systematically capture substrate, water intercalation, and graphene layer effects (Fig.~\ref{fig1}a), we constructed a series of atomistic models and adopted a consistent naming scheme. Each system is labeled as $S_nL_m$, where $n$ denotes the number of intercalated water molecules in our simulation cell and $m$ indicates the number of graphene layers. The water slab, which is present in all models, is not explicitly included in the naming scheme. If $S$, $n$, or $L_m$ is omitted in the name, it implies the absence of the substrate, intercalated water, or graphene layers, respectively. According to this convention:

• $L_1$ and $L_5$ denote systems comprising one- and five-layer graphene, respectively, without substrate or intercalated water, and with a water slab placed above the graphene.

• $S$ refers to the bare \ce{CaF2} substrate with a water slab.

• $SL_1$, $SL_4$, and $SL_5$ denote systems consisting of the substrate (without intercalated water), one-, four-, and five-layer graphene, respectively, together with an overlying water slab.

• $S_nL_1$ and $S_nL_4$ denote systems with one- and four-layer graphene, respectively, containing $n$ intercalated water molecules between the graphene and substrate, along with an overlying water slab.

In the work by Li et al.\cite{RN528}, the binding energies of different phases of monolayer ice and monolayer amorphous water confined between monolayer graphene and the \ce{CaF2} substrate were compared. Their results showed that one particular ice phase was more stable than the other configurations. Motivated by their findings, we extended our study beyond models with discrete numbers of intercalated water molecules to include several ordered monolayer ice structures, as well as one double-layer amorphous water confined between the substrate and monolayer graphene. The specific structural configurations are illustrated in Fig. S9. The naming convention for these systems is as follows:

• $S_{\mathrm{H}}L_1$: a monolayer of hexagonal ice (ice-Ih).

• $S_{\mathrm{P}}L_1$: a monolayer of pentagonal ice.

• $S_{\mathrm{S1}}L_1$ and $S_{\mathrm{S2}}L_1$: two different monolayer square ice structures.

• $S_{\mathrm{D}}L_1$: a double-layer amorphous water configuration.

This naming scheme enables a systematic comparison across different structural and interfacial configurations.

%\subsection{Training set data generation and potential training}
\subsection{Potential training and accuracy}

To train the MLIP, a representative dataset was first constructed using on-the-fly machine learning force field MD simulations.\cite{PhysRevLett.122.225701, PhysRevB.100.014105, 10.1063/5.0009491} This approach significantly accelerates MD simulations for complex systems such as $SL_5$ and $S_nL_4$. Atomic configurations, along with their corresponding DFT-computed energies and forces, were extracted from the MD trajectories to form the initial dataset. The ACE method\cite{RN4, RN22, lysogorskiy2021performant} was employed to train MLIPs. Given the structural complexity and diversity of systems involved, as described in Section~\ref{subsec2.1}, it is impractical to develop one potential that performs accurately across all configurations. Therefore, multiple system-specific potentials were trained: $ACE-L_1$ for $L_1$, $ACE-L_5$ for $L_5$, $ACE-S$ for $S$, $ACE-SL_1$ for $SL_1$, $ACE-SL_4$ for $SL_4$, $ACE-SL_5$ for $SL_5$, $ACE-S_nL_1$ for all monolayer graphene intercalated water systems, and $ACE-S_nL_4$ for all four-layer graphene intercalated water systems. To ensure the reliability of each potential for large-scale and long-timescale MD simulations, we further refined MLIPs using an active learning (AL) strategy.\cite{PhysRevMaterials.7.043801} The workflow of the refinement process is illustrated in Fig.~\ref{fig7}a. Detailed simulation and training parameters are provided in Section~\ref{Methods}. The precision of the trained potentials was evaluated using the root mean square error (RMSE) of energies and forces on the test set, as summarized in Table S1. All trained potentials exhibit energy RMSE values no more than 0.50 meV/atom and force RMSE values below 40 meV/Å, indicating high accuracy of our MLIPs. The time evolution of the temperature of all water molecules, along with the total internal energy of each system during the NVE ensemble MD simulations, is presented in the Supplementary Information. Throughout the NVE ensemble MD simulations, the total energy of each system fluctuates by approximately 0.1 eV, further confirming the stability and reliability of the trained MLIPs.

%\subsection{Graphene layer stacking effect}
\subsection{Effect of various graphene layers} % on its vSFG spectra}

\begin{figure}[ht]
\centering
\includegraphics[width=1.0\textwidth]{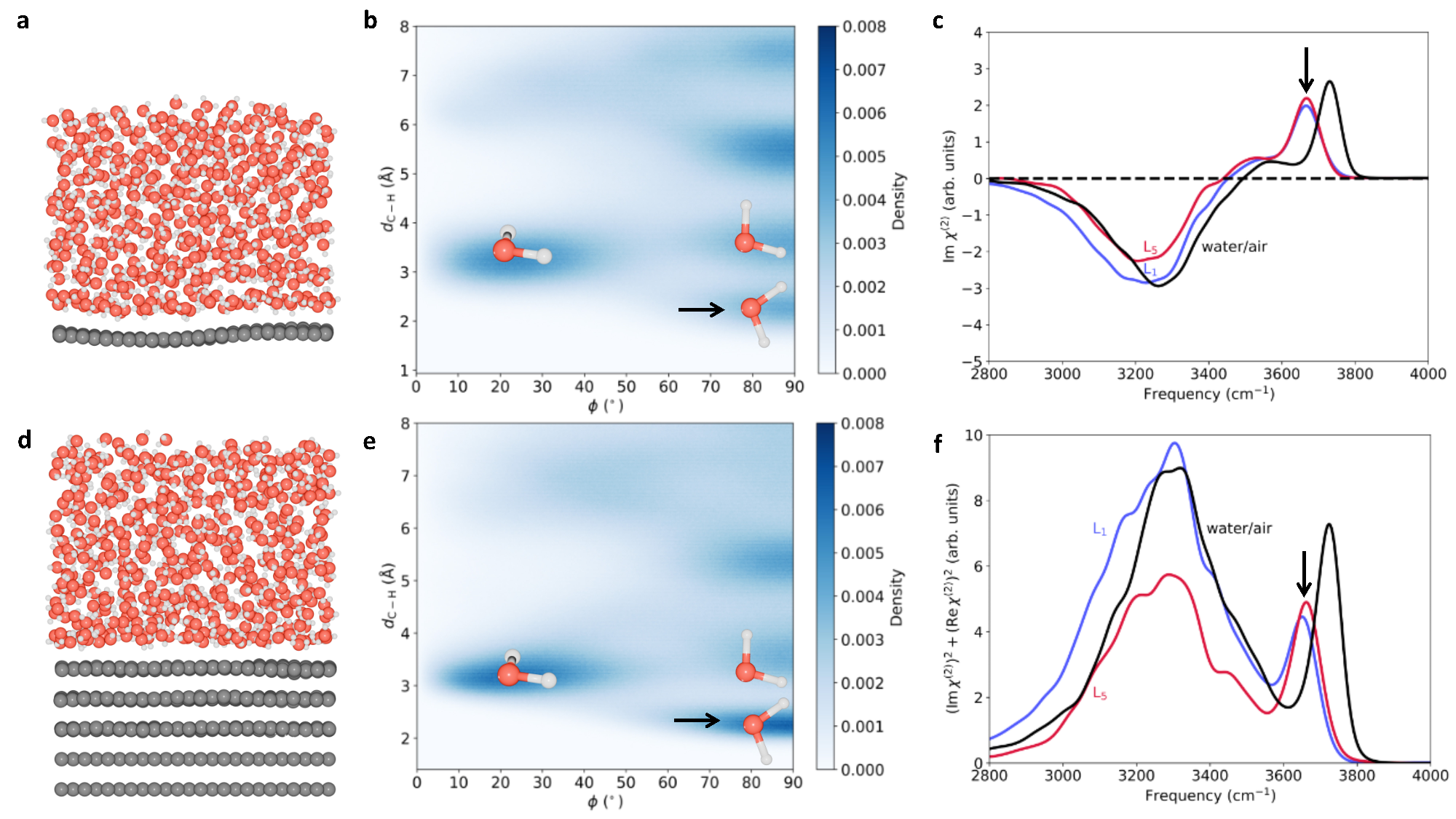}
\caption{\textbf{a} and \textbf{d}, Atomistic models of the $L_1$ and $L_5$ systems, respectively. \textbf{b} and \textbf{e}, Angle distribution in the first water layer adjacent to graphene in $L_1$ and $L_5$, respectively. For direct comparison, only the distribution of the angle $\phi$ is shown here. The corresponding angle $\sigma$ distributions are provided in the Supplementary Information(Fig. S30). \textbf{c}, Imaginary part of the vSFG spectra, Im($\chi^{(2)}$), for $L_1$, $L_5$ and water-air interface. The black dashed line indicates the zero reference level in the figure. \textbf{f}, The square modulus $|\chi^{(2)}|^2$ of the vSFG spectra (sum of the squared imaginary and real parts) for $L_1$, $L_5$ and water-air interface. The black arrow highlights the correspondence between the interfacial water molecular configuration and its characteristic peak in the vSFG spectra.}
\label{fig2}
\end{figure}

It is widely accepted that graphite exhibits hydrophobic properties. However, the wettability of monolayer and multilayer graphene remains a topic of debate.\cite{BELYAEVA2020100482} In this study, we first considered monolayer ($L_1$) and five-layer graphene ($L_5$) systems without substrates and defects. To analyze the orientation of interfacial water molecules, we defined two angles (Fig.~\ref{fig1}b): $\sigma$, which describes the orientation of the lower coordination O-H bond with respect to the Z-axis, and $\phi$, representing the angle between the normal vector of the water plane and the Z-axis. As shown in Fig.~\ref{fig2}b and~\ref{fig2}e, for the first water layer adjacent to graphene, both $L_1$ and $L_5$ systems exhibit three distinct types of water molecule orientations. Among them, the water molecules closest to the graphene surface often have one O-H bond pointing directly toward the surface (also known as dangling O-H). Notably, the $L_5$ system shows a significantly higher number of such dangling O-H bonds compared to the $L_1$ system. This suggests that $L_5$ is more hydrophobic than $L_1$, consistent with experimental measurements of contact angles, which showed an increase in hydrophobicity with the number of graphene layers.\cite{RN91, RN496, munz2015thickness} In contrast, the angular distribution at the water-air interface does not exhibit such well-defined orientation types(Fig. S29), reflecting the weaker constraints imposed by the gas-liquid interface compared with the stronger confinement at the solid–liquid interface.\cite{wang2025spectralsimilaritymasksstructural}

We then calculated the vSFG spectra for $L_1$ and $L_5$ systems, using the well-studied water-air interface as a reference. Experimentally, the vSFG peak corresponding to the dangling O–H stretch of water-air interface appears at approximately 3700 cm$^{-1}$.\cite{RN665, doi:10.1073/pnas.1906243117, 10.1063/1.3135147, RN666, RN669} As shown in Fig.~\ref {fig2}c, the simulated imaginary part, Im($\chi^{(2)}$), of the vSFG spectra of water-air interface shows the dangling O–H peak at 3730 cm$^{-1}$. The slight blue shift relative to experiment value is attributed to the neglect of nuclear quantum effects in our MD simulations.\cite{RN442, annurev:/content/journals/10.1146/annurev-physchem-090722-124705} For $L_1$, the simulated vSFG spectrum agrees well with both previous simulations and experimental measurements,\cite{RN493, wang2025spectralsimilaritymasksstructural, du2025machinelearningacceleratedcomputational, RN669} confirming its hydrophobic nature. The dangling O-H peak of $L_1$ is red-shifted by 65 cm$^{-1}$ relative to the water-air interface, which is attributed to O-H-$\pi$ interactions that weaken the O-H bond and lower its vibrational frequency.\cite{RN62, RN91, RN669} For $L_5$, water molecules close to the graphene surface have a higher ratio of dangling O-H bonds and consequently less hydrogen-bonded O-H. This molecular arrangement is reflected in Im($\chi^{(2)}$) spectra. Compared to Im($\chi^{(2)}$) spectrum of $L_1$, Im($\chi^{(2)}$) spectrum of $L_5$ shows a reduced amplitude around 3200 cm$^{-1}$ (associated with hydrogen-bonded O-H stretch vibrations) and an enhanced amplitude near 3665 cm$^{-1}$(characteristic of the dangling O-H groups). These differences are more obvious in the square modulus $|\chi^{(2)}|^2$ of the vSFG spectra (sum of the squared imaginary and real parts)(Fig.~\ref {fig2}f). These findings further confirm that both monolayer and multilayer pristine graphene are intrinsically hydrophobic, with the degree of hydrophobicity increasing with the number of layers. This trend is consistent with the enhanced exposure of the dangling O-H groups and is supported by experimental contact angle measurements.\cite{RN91, RN496, munz2015thickness}

\subsection{Substrate effect}

\begin{figure}[ht]
\centering
\includegraphics[width=1.0\textwidth]{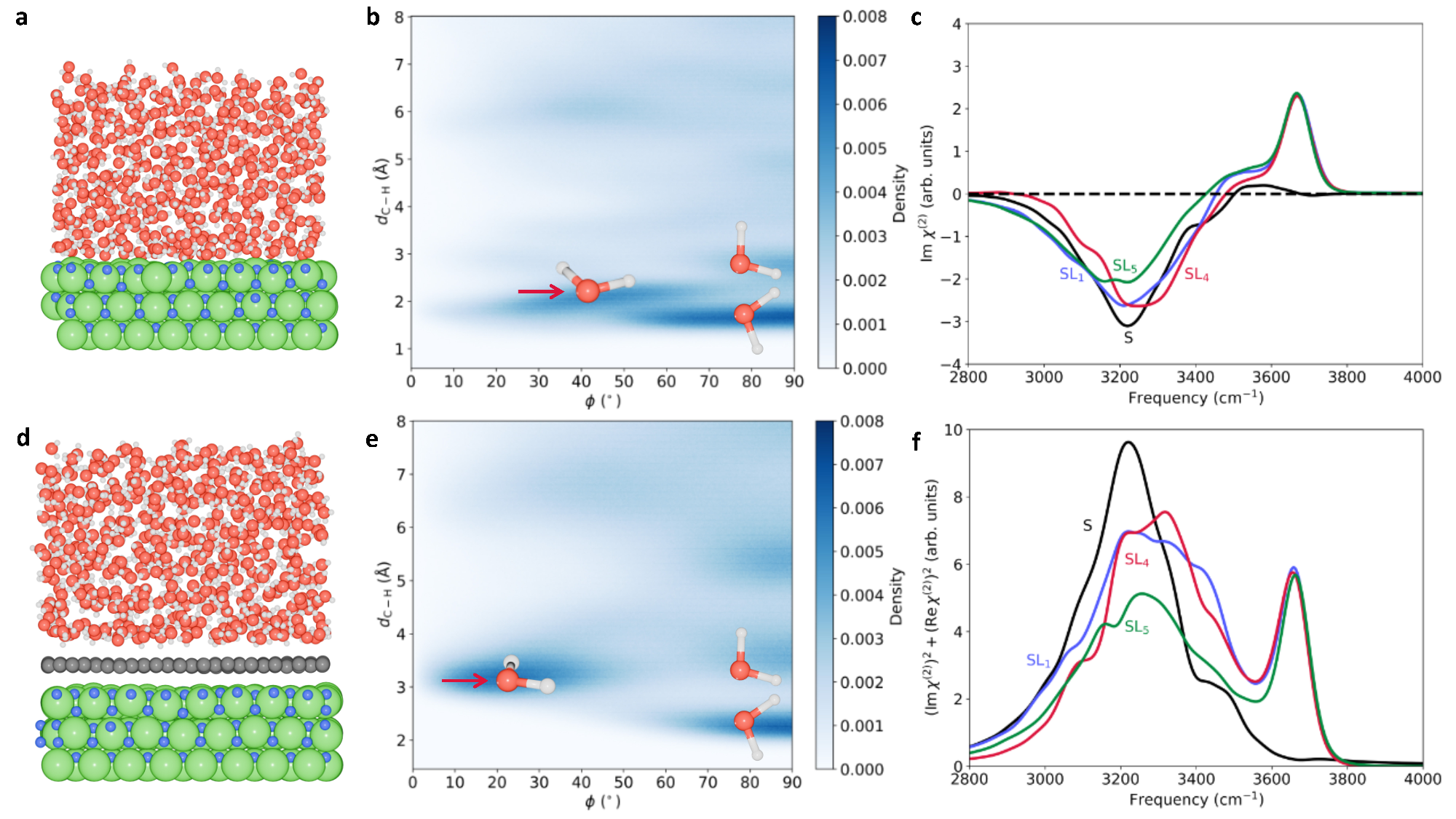}
\caption{\textbf{a} and \textbf{d}, Atomistic models of the $S$ and $SL_1$ systems, respectively. \textbf{b} and \textbf{e}, Angle distribution of the first water layer adjacent to the substrate or graphene for $S$ and $SL_1$ systems, respectively. For direct comparison, only the distribution of the angle $\phi$ is shown here. The corresponding angle $\sigma$ distributions are provided in the Supplementary Information(Fig. S31). \textbf{c}, Imaginary part of the vSFG spectra, Im($\chi^{(2)}$), for $S$, $SL_1$, $SL_4$ and $SL_5$. The black dashed line indicates the zero reference level in the figure. \textbf{f}, The square modulus $|\chi^{(2)}|^2$ of the vSFG spectra (sum of the squared imaginary and real parts) for $S$, $SL_1$, $SL_4$ and $SL_5$.}
\label{fig3}
\end{figure}

Most experimental studies reporting vSFG spectra of graphene used \ce{CaF2} as a substrate, owing to its wide optical transparency range and the chemically inert nature of its (111) surface.\cite{RN97, RN103, RN91, RN75, RN650, RN591, RN63, RN62, RN60, RN662} The vSFG spectrum of monolayer graphene on the \ce{CaF2} substrate often exhibits features indicative of hydrophilic behavior.\cite{RN662, RN97, RN91} Since \ce{CaF2} is well known to be hydrophilic and has a strong affinity for water molecules,\cite{RN522, RN651, RN525, RN652} this observation appears to contradict the intrinsic hydrophobicity of pristine monolayer graphene, as confirmed by both simulations and experimental measurements.\cite{RN52, RN493} To explain this apparent contradiction, some investigators have proposed the concept of wetting transparency,\cite{RN499, RN496} where the wetting properties of the underlying substrate (\ce{CaF2}) dominate the observed interfacial behavior despite the presence of the graphene layer. To investigate the substrate effect, we computed water orientation with respect to the substrate or graphene surface as well as vSFG spectra for the following systems: (i) the \ce{CaF2} substrate and a 512 water molecule slab (S) (Fig.~\ref {fig3}a), (ii) the \ce{CaF2} substrate with monolayer graphene and the water slab ($SL_1$)(Fig.~\ref {fig3}d), (iii) the \ce{CaF2} substrate with four-layer graphene and the water slab ($SL_4$)(Fig. S32a), and (iv) the \ce{CaF2} substrate with five-layer graphene and the water slab ($SL_5$)(Fig. S32d).

Based on the atomic distribution analysis of system $S$ (Fig.~\ref {fig3}b), the average position of the first water layer is closer to \ce{CaF2} surface than to the graphene surface ($SL_1$). This proximity is attributed to the stronger interactions between water oxygen atoms and calcium ions on the \ce{CaF2} surface, in contrast to the relatively weak physisorption observed between water molecules and graphene. As a result, the fraction of dangling O-H bonds near the \ce{CaF2} surface is significantly reduced, reaching only about 60\% of that observed in the $L_1$ or $L_5$ systems. Detailed information on the calculation of dangling O-H bonds is provided in the Supplementary Information. In contrast to graphene-based systems($L_1$ and $L_5$), the second type of interfacial water in $S$ (indicated by the red arrow) is oriented toward the bulk water rather than lying parallel to the surface. This distinct interfacial arrangement leads to a markedly different Im($\chi^{(2)}$) spectral profile (Fig.~\ref {fig3}c). Unlike graphene systems ($L_1$ and $L_5$), the \ce{CaF2} and water interface does not exhibit a positive peak in the Im($\chi^{(2)}$) spectra around 3665 cm$^{-1}$, which in vSFG spectra is indicative of dangling O–H groups. Given the sensitivity of vSFG to molecular orientation, the presence of even a small number of dangling O-H bonds would generally produce this positive peak. The absence of such a feature in $S$ can be attributed to two factors: (1) the significantly reduced number of dangling O-H bonds, and (2) partial cancellation of the dangling O-H signal by the non-hydrogen-bonded, second-type water molecules oriented toward the bulk. Moreover, our simulated vSFG spectrum for system $S$ shows good agreement with the simulation and experimental measurements,\cite{RN97, RN60, RN91} showing only a single broad negative peak characteristic of hydrogen-bonded O-H. This consistency supports the validity of our structural model and interpretation.

The angle distribution of water molecules in $SL_1$(Fig.~\ref {fig3}e), $SL_4$(Fig. S32b), and $SL_5$(Fig. S32e) systems is quite similar, and closely resembles that of the corresponding systems without a substrate ($L_1$ and $L_5$). In the first water layer adjacent to the graphene surface, three distinct configurations of water molecules are observed. Among these, the water molecules closest to the graphene typically adopt a configuration in which one O–H bond points toward the graphene surface. In the substrate-supported systems ($SL_1$, $SL_4$, and $SL_5$), the number of dangling O-H bonds increases with the number of graphene layers and is consistently higher than the corresponding substrate-free systems ($L_1$ and $L_5$). The elevated number of dangling O–H bonds in the first water layer of substrate-supported systems suggests that the presence of the \ce{CaF2} substrate may enhance the hydrophobic character of the interface. This observation contrasts with the concept of wetting transparency. 

In addition to the similar angular distributions of water molecules observed in systems $L_1$, $L_5$, $SL_1$, $SL_4$, and $SL_5$, Im($\chi^{(2)}$) spectra exhibit a generally consistent spectral profile across all these five systems. Although there are minor differences in the intensities of the positive and negative peaks, attributable to differing amounts of hydrogen-bonded versus dangling O-H groups, the overall shape of the spectra remains largely unchanged. Furthermore, the square modulus of the vSFG spectra, $|\chi^{(2)}|^2$ (Fig.~\ref {fig3}f), derived from our simulations, differs from the experimental spectra reported by Kim et al. \cite{RN91}. Notably, no significant change is observed at the frequency of 3665 cm$^{-1}$ as the number of graphene layers increases. These simulation results provide no evidence supporting wetting transparency in monolayer and multilayer graphene systems, suggesting that the substrate alone is insufficient to induce a substantial change in the vSFG spectral response.

\subsection{Intercalated water effect}

To explain the deviations between our simulated and experimental vSFG results for supported multilayer graphene, we first consider the possible importance of water intercalation. Experimentally, no dangling O-H peaks are observed in the vSFG spectra when up to three stacked graphene layers.\cite{RN91} In contrast, our simulations for the monolayer graphene system exhibit a pronounced dangling O-H peak(Fig.~\ref {fig3}f), suggesting that other factors such as intercalated water may significantly affect the spectral features. It is well established that a thin layer of water readily forms on hydrophilic surfaces under ambient conditions.\cite{C5SM02143J, RN656, 10.1115/1.2826087, li2015two} When graphene is transferred onto a hydrophilic substrate, water molecules can be trapped beneath the impermeable graphene layer.\cite{dollekamp2017charge, okmi2022discovery} Furthermore, due to the relatively low migration barrier (approximately 0.19 eV) and high diffusion coefficient (8.35 × 10$^{-6}$ cm$^{2}$/s) of water,\cite{RN528} it is highly likely to have intercalated water molecules between the hydrophilic substrate and graphene. Additionally, during vSFG spectral measurements, where samples are typically immersed in water,\cite{RN91, RN97, RN62, RN591} capillary effect can further promote the intercalation of water molecules between the graphene and hydrophilic substrate interface. We thus explicitly consider the influence of varying amounts of intercalated water molecules on the vSFG spectra.

\begin{figure}[ht]
\centering
\includegraphics[width=1.0\textwidth]{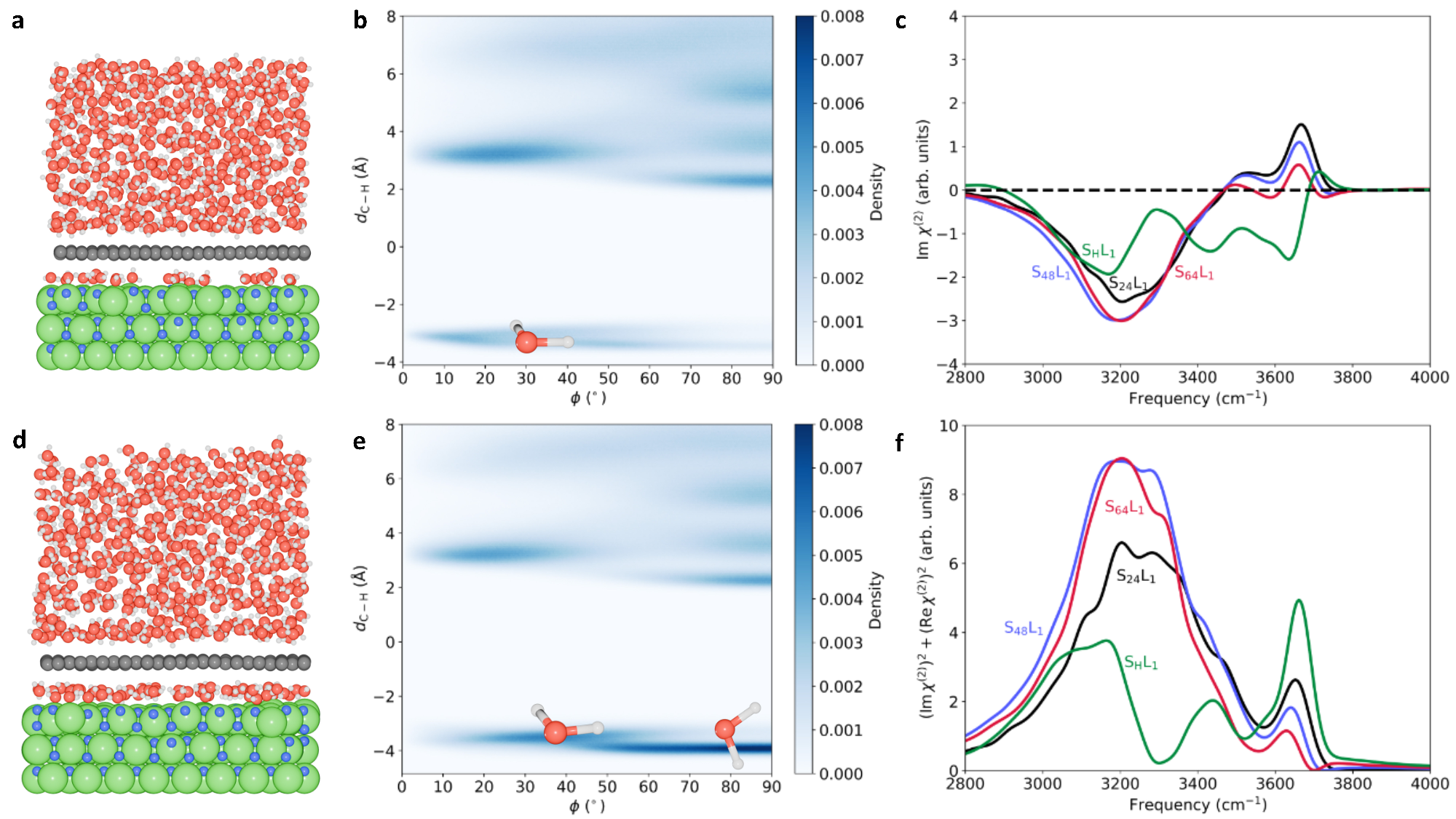}
\caption{\textbf{a} and \textbf{d}, Atomistic models of the $S_{24}L_1$ and $S_{64}L_1$ systems, respectively. \textbf{b} and \textbf{e}, Angle distribution of the intercalated water for $S_{24}L_1$ and $S_{64}L_1$ systems, respectively. For direct comparison, only the distribution of the angle $\phi$ is shown here. The corresponding angle $\sigma$ distributions are provided in the Supplementary Information(Fig. S33b and S34b). \textbf{c}, Imaginary part of the vSFG spectra, Im($\chi^{(2)}$), for $S_{24}L_1$, $S_{48}L_1$, $S_{64}L_1$ and $S_{\mathrm{H}}L_1$. The black dashed line indicates the zero reference level in the figure. \textbf{f}, The square modulus $|\chi^{(2)}|^2$ of the vSFG spectra (sum of the squared imaginary and real parts) for $S_{24}L_1$, $S_{48}L_1$, $S_{64}L_1$ and $S_{\mathrm{H}}L_1$.}
\label{fig4}
\end{figure}

In our model, a full monolayer of intercalated ice-Ih between the \ce{CaF2} substrate and monolayer graphene consists of 96 water molecules. In addition to this full monolayer, we also considered systems with partial coverage corresponding to one-quarter, one-half, two-thirds, and five-sixths of a monolayer of ice-Ih, containing 24, 48, 64, and 80 water molecules, respectively. These systems are denoted as $S_{24}L_1$(Fig.~\ref {fig4}a), $S_{48}L_1$(Fig. S33d), $S_{64}L_1$(Fig.~\ref {fig4}d) and $S_{80}L_1$(Fig. S34d), respectively. The system with a complete ice-Ih monolayer is labeled $S_{\mathrm{H}}L_1$, as the water molecules form hexagonal rings (hexamers)(Fig. S9a) stabilized by hydrogen bonding, characteristic of the ice-Ih structure. Beyond the monolayer ice-Ih configuration, we also explored alternative initial interfacial water structures, including different phases of monolayer ice and a bilayer of amorphous water, to include the possibility of metastable water structures that could influence wettability. The configuration of these water structures is shown in Fig. S9. 

We first considered systems containing intercalated water molecules in quantities equal to or less than that of a complete monolayer of ice-Ih. We generally found most of the intercalated water lying flat on the substrate for small water coverage ($S_{24}L_1$, Fig.~\ref{fig4}a). When water coverage approached a full monolayer, water was increasingly oriented with O-H towards the substrate (Fig.~\ref {fig4}e), becoming the dominating configuration at $S_{80}L_1$ (Fig. S34f). For a complete monolayer coverage ($S_{\mathrm{H}}L_1$)(Fig. S35a), additionally, one more water configuration with O-H pointing towards the graphene surface is found. %The orientation of the water molecules in the water slab above the graphene layer was merely unaffected by the intercalated water. 

It is noteworthy that the ideal hexagonal ring pattern characteristic of crystalline ice-Ih is not preserved after equilibration in our simulations (Fig. S40g). This observation aligns with experimental observations from room-temperature atomic force microscopy, which also report the absence of well-ordered ice-Ih under similar conditions.\cite{RN526} Specifically, while ice-Ih has been detected between graphene and substrates \ce{BaF2}(111), it is not observed between graphene and substrate \ce{CaF2}(111). This difference is primarily attributed to the lattice mismatch between the \ce{CaF2}(111) surface and the crystalline structure of ice-Ih. These results suggest that the orientation and structure of intercalated water molecules are mainly governed by the degree of water coverage rather than a specific structural arrangement. It is also important to note that at sub-monolayer coverages, water does not form a homogeneous structure but instead organizes into discrete, localized clusters(Fig. S40d, S40e and S40f).

We then calculated vSFG spectra as a function of intercalated water coverage. As shown in Fig.~\ref {fig4}c and Fig. S40a, the amplitude of the positive peak near 3665 cm$^{-1}$ in the Im($\chi^{(2)}$) spectra gradually decreases as the number of intercalated water molecules increases. In the $SL_1$ system (without intercalated water), the Im($\chi^{(2)}$) spectra display a shoulder around 3500 cm$^{-1}$ (Fig.~\ref {fig3}c). This shoulder originates from the antisymmetric O-H stretch mode of interfacial water molecules with two donor hydrogen bonds.\cite{stiopkin2011hydrogen} As the amount of intercalated water molecules increases, the shoulder evolves into a distinct peak, which shifts to more negative values. As shown in Fig. S40c, if the Im($\chi^{(2)}$) signal is restricted to only the intercalated water molecules, it exhibits a negative peak near 3650 cm$^{-1}$ and a positive peak around 3285 cm$^{-1}$. The amplitude of negative peaks increases with the increase of intercalated water molecules. This trend is primarily due to the growing number of water molecules with one O-H pointing toward the graphene, which contributes strongly to the sharp negative peak at 3650 cm$^{-1}$ in the $S_{\mathrm{H}}L_1$ system. Furthermore, in $S_{\mathrm{H}}L_1$, contributions from water molecules on both sides of graphene lead to a more complex Im($\chi^{(2)}$) spectrum with three negative peaks(Fig.~\ref {fig4}c). In the square modulus $|\chi^{(2)}|^2$ of the vSFG spectra (Fig.~\ref {fig4}f), the intensity of the peak at 3650 cm$^{-1}$ first decreases and then increases with the addition of intercalated water. These findings suggest that the experimentally observed vSFG spectra indicating hydrophilic behavior of monolayer graphene on \ce{CaF2} substrates, despite the intrinsic hydrophobicity of pristine graphene, may be attributed to the presence of intercalated water molecules. During vSFG measurements, where samples are immersed in water, capillary forces can drive water molecules beneath the graphene, thereby altering the interfacial water structure and vSFG response.

To gain deeper insight into how intercalated water structure and coverage affect the vSFG response, we further explored a range of crystalline monolayer ice models along with a representative double-layer amorphous water configuration. Across all monolayer ice systems, the initial crystalline order was lost during equilibration(Fig. S41). Importantly, similar water orientations and vSFG spectral characteristics were observed in systems with identical surface coverage(Fig. S41), independent of their initial structure. In contrast, the double-layer amorphous water configuration exhibited distinct spectral signatures, characterized by vSFG features. These results indicate that the vSFG response of intercalated water is more strongly influenced by intercalated water surface coverage than by the initial ice phase. Detailed structural and spectral analyses are provided in the Supplementary Information.

\begin{figure}[ht]
\centering
\includegraphics[width=1.0\textwidth]{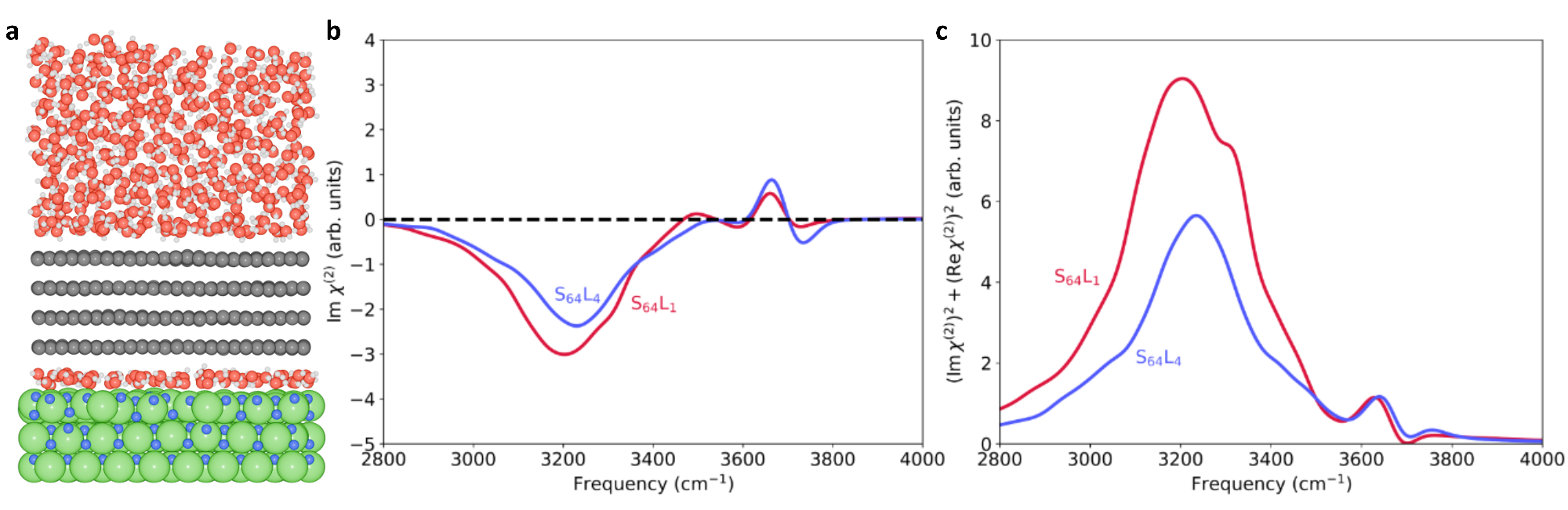}
\caption{\textbf{a}, Atomistic model of the $S_{64}L_4$ system. \textbf{b}, Imaginary part of the vSFG spectra, Im($\chi^{(2)}$), for $S_{64}L_1$and $S_{64}L_4$. The black dashed line indicates the zero reference level in the figure. \textbf{c}, The square modulus $|\chi^{(2)}|^2$ of the vSFG spectra (sum of the squared imaginary and real parts) for $S_{64}L_1$ and $S_{64}L_4$.}
\label{fig5}
\end{figure}

In addition to investigating monolayer graphene with intercalated water molecules, we further examined the influence of graphene thickness on the behavior of intercalated water molecules and the corresponding vSFG spectra. Specifically, we selected a four-layer graphene system as the representative case to analyze the angular distribution and vSFG spectral features. The simulated systems consist of a \ce{CaF2} substrate, a four-layer graphene sheet, a water slab containing 512 water molecules, and varying amounts of intercalated water (24, 48, 64, and 96 molecules). These systems are referred to as $S_{24}L_4$(Fig. S38a), $S_{48}L_4$(Fig. S38d), $S_{64}L_4$(Fig.~\ref {fig5}a), and $S_{\mathrm{H}}L_4$(Fig. S39d), respectively. 

The angle distribution of intercalated water molecules between the \ce{CaF2} substrate and four-layer graphene exhibited a trend similar to that observed in systems with monolayer graphene. Specifically, as the amount of intercalated water increases, the orientation of water molecules evolves in a comparable manner to that of monolayer graphene. In Im($\chi^{(2)}$) (Fig.~\ref {fig5}b), taking $S_{64}L_4$ as an example, compared to the monolayer graphene systems($S_{64}L_1$), the broad peak at the frequency of 2800-3500 cm$^{-1}$ exhibits a blue shift. Additionally, the sharp peak near 3665 shows a high amplitude in the four-layer graphene system. For thicker graphene layers, the reduced number of hydrogen-bonded O–H groups and the increased presence of dangling O–H bonds at the interface jointly lead to an interfacial arrangement consistent with a more hydrophobic character. The square modulus vSFG spectra, $|\chi^{(2)}|^2$ (Fig.~\ref {fig5}c), also follow a trend similar to that of the monolayer systems, but with a blue-shifted broad peak and an unobvious peak around 3650 cm$^{-1}$. These spectral features further confirm the stronger hydrophobic character of multilayer graphene and its impact on the interfacial water structure.

\subsection{Driving force for water intercalation}

\begin{figure}[ht]
\centering
\includegraphics[width=1.0\textwidth]{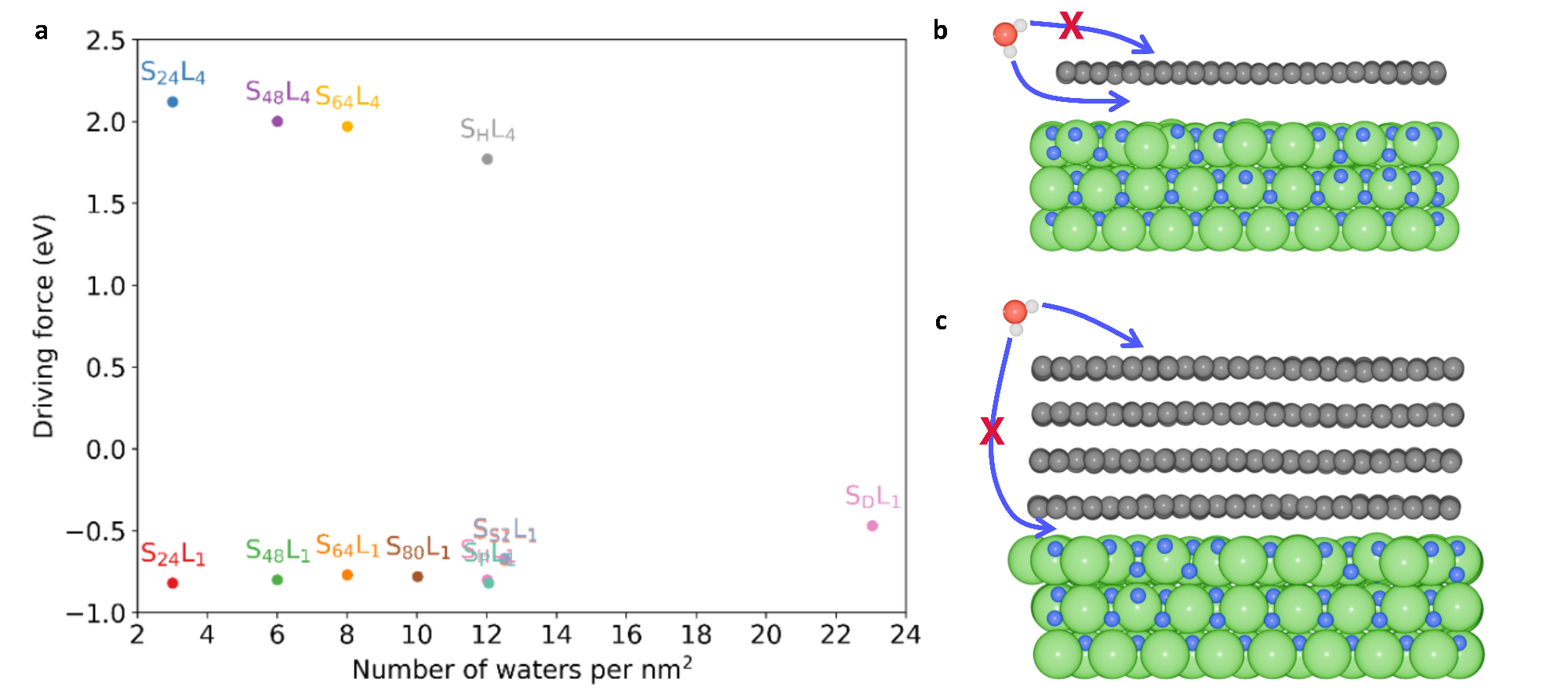}
\caption{\textbf{a}, Free energy of intercalated water in monolayer and four-layer graphene systems at different coverages.
The free energy per water molecule in the corresponding water slab is set as the reference (zero). The plot shows the relative free energy of each intercalated water molecule compared to the reference. \textbf{b}, Schematic illustration showing that water molecules energetically prefer intercalation between monolayer graphene and substrate over surface adsorption. \textbf{c}, Schematic illustration showing that water molecules energetically prefer surface adsorption on four-layer graphene over intercalation. }
\label{fig6}
\end{figure}

Although the same amount of intercalated water molecules yields similar trends in the vSFG spectra for both monolayer and four-layer graphene systems (Fig.~\ref {fig5}c), experimental vSFG measurements reveal notable differences between them.\cite{RN91} Specifically, for graphene films thicker than three layers, a distinct peak appears in the frequency range of 3600-3700 cm$^{-1}$, while this feature is absent in systems with one to three graphene layers.

To understand this discrepancy, we calculated the free energy of intercalated water molecules and water molecules in the slab along the trajectory from internal energy data and an additional entropic contribution (see Section~\ref{entropylabel} for details). We used the free energy per water molecule in the slab as the reference, setting its value to zero (Table S6). Our results show that the free energy of intercalated water molecules is negative in the monolayer graphene system (Fig.~\ref {fig6}a), indicating that water intercalation between graphene and the \ce{CaF2} substrate is thermodynamically favorable. Furthermore, the free energy remains nearly unchanged as the number of intercalated water molecules increases up to 12 per nm$^2$($S_{\mathrm{H}}L_4$ and $S_{\mathrm{P}}L_4$). Beyond this coverage, further intercalation becomes energetically less favorable, as observed in systems such as $S_{\mathrm{S1}}L_1$ (12.5 water/nm$^2$), $S_{\mathrm{S2}}L_1$ (12.5 water/nm$^2$), and especially $S_{\mathrm{D}}L_1$ (23 water/nm$^2$). In contrast, the free energy of an intercalated water molecule becomes positive ($>$ 1.5 eV) for the four-layer graphene system (Fig.~\ref {fig6}a), indicating an energetically unfavorable intercalation process. However, with increasing numbers of intercalated water molecules, the free energy becomes progressively less unfavorable, likely due to the formation of hydrogen bonds that partially stabilize the confined water structure. These findings imply that, during vSFG measurements where the samples are immersed in water, water molecules are likely to be intercalated under monolayer graphene to minimize the system’s free energy (Fig.~\ref {fig6}b). However, for thicker graphene ($n \ge 4$), water intercalation is energetically disfavored and thus unlikely to occur (Fig.~\ref {fig6}c). This provides a plausible explanation for the experimentally observed hydrophilic vSFG features of monolayer graphene on a \ce{CaF2} substrate. Rather than resulting from wetting transparency, these spectral signatures are more likely attributable to the presence of intercalated water, which modifies the interfacial structure and vibrational response, yielding features typically associated with hydrophilic surfaces.

\section{Discussion}

Our results reveal that monolayer graphene supported on hydrophilic substrates, such as \ce{CaF2}, readily permits environmental water molecules to intercalate spontaneously at the graphene-substrate interface. This intercalation may substantially modify the interfacial structure, thereby exerting a pronounced influence on the physical properties of the system. Notably, experimental studies have shown that the electronic bandgap of graphene can be tuned from 0.029 eV up to 0.2 eV by modulating the amount of adsorbed water.\cite{RN649} These adsorbed waters also degrade carrier mobility and shift the Dirac voltage in graphene field-effect transistors, primarily due to charged impurity scattering.\cite{RN664} Hence, during fabrication of van der Waals heterostructures, graphene-coated metals, or graphene-based devices (e.g., field-effect transistors on \ce{CaF2}), careful control over environmental water is critical to avoid unintended intercalation that may adversely affect device performance and stability. 

Our theoretical analysis also calls for a reassessment of vSFG spectra interpretation for graphene on hydrophilic substrates. Although thermal treatments are commonly applied to remove pre-existing intercalated water before measurement,\cite{RN91}, the immersion required for vSFG experiments facilitates facile water diffusion beneath the graphene through capillary forces. To suppress such spontaneous water ingress during vSFG measurements, graphene monolayers should be transferred under high-vacuum conditions onto pristine substrates, and graphene edges physically sealed—for example, by thin gold films deposition\cite{wang2025spontaneoussurfacechargingjanus}—to effectively block water entry. This protocol is essential for accurately probing graphene’s intrinsic wetting behavior and clarifying any genuine wetting transparency effects.

Furthermore, our findings highlight that the extent of intercalated water coverage critically shapes the vSFG spectral response. Operando vSFG spectroscopy measurements performed under controlled and varying intercalation conditions could provide valuable insights into the intricate relationship between interfacial water structuring and corresponding spectral features. Notably, Wang et al.\cite{RN662} investigated the effect of intercalated water content on the vSFG response by varying ambient humidity, which modulates the amount of water trapped between monolayer graphene and the \ce{CaF2} substrate. However, their study did not consider water molecules on the opposite side of the graphene. Our simulations demonstrate that the coexistence of intercalated and adsorbed water layers induces complex spectral interplay, as evident in the Im($\chi^{(2)})$ spectra of the $S_{\mathrm{H}}L_1$ system (Fig.~\ref{fig4}c). This dual-interfacial configuration scenario is worth investigating experimentally, as it may better reflect realistic conditions and improve the understanding of graphene–water interfacial behavior.

In our present work, we only consider substrates without defects. For graphene on a pristine \ce{CaF2}(111) surface, our simulation results indicate non-wetting transparency. However, in practice, \ce{CaF2} substrates may contain defects, as fluorine ions can dissociate from the surface during sample preparation, especially when the substrate is treated in a strong acidic solution to produce a fresh \ce{CaF2} surface.\cite{RN62} Experimentally reported surface charge densities for the \ce{CaF2}(111) surface are 21.5 $\mathrm{mC/m^2}$\cite{RN62} and 40 $\mathrm{mC/m^2}$\cite{RN662}, respectively. In the theoretical work of Wehling et al.,\cite{10.1063/1.3033202} it was reported that for monolayer graphene on passivated \ce{SiO2}, no additional bands appear at the Fermi level beyond graphene’s Dirac bands, and no doping occurs. In contrast, for graphene on defective \ce{SiO2}, the dipole moments of adsorbed water molecules can generate local electrostatic fields, shifting the substrate’s defect states relative to the graphene electrons and altering their hybridization with graphene’s bands. Such effects may also play an important role in the vSFG spectra and are likely relevant for estimating graphene’s wetting transparency with respect to the substrate. To accurately capture the influence of defects on graphene and interfacial water, it is necessary to include long-range electrostatic interactions. Our current machine learning model considers only short-range interaction, which is insufficient for describing defect-related effects. We will address the influence of substrate defects in future work.

\section{Conclusion}

In this theoretical study, we have systematically investigated the effects of stacked graphene layers, substrate, and the presence of intercalated water molecules on the interfacial behavior of graphene-water systems using machine-learning-assisted MD simulations and vSFG spectra simulations. Our results indicate that graphene is non-wetting transparent to the defect-free hydrophilic substrate. The hydrophilic features observed in experimental vSFG spectra for supported monolayer graphene are primarily attributed to the combined contribution of water molecules located on both sides of graphene, particularly due to the spontaneous intercalation of water between monolayer graphene and the hydrophilic \ce{CaF2} substrate. We confirm that pristine graphene is hydrophobic, and the hydrophobicity becomes more pronounced as the number of graphene layers increases. While monolayer graphene readily allows for water intercalation, either during preparation under ambient humidity or upon immersion in water during vSFG spectra measurements, this process becomes energetically unfavorable for multilayer graphene, e.g., four layers and more, making intercalation less likely to occur under the same conditions. 

Our work provides a mechanistic, atomistic-level explanation for the distinct variations observed in the vSFG spectra between monolayer and multilayer graphene systems. Furthermore, these insights offer important guidance for the design and fabrication of van der Waals heterostructures and graphene-based devices, particularly in managing environmental conditions to prevent unintentional water intercalation, which can adversely affect device performance and complicate interfacial characterization.

\section{Methods}\label{Methods}

\subsection{Reference data}

To parametrize MLIPs using the ACE method \cite{RN4, RN22}, one needs to perform DFT calculations of total energies and atomic forces for a sufficiently large number of representative reference configurations. These reference structures were generated using the on-the-fly machine-learning force field method\cite{PhysRevLett.122.225701, PhysRevB.100.014105, 10.1063/5.0009491} as implemented in VASP6.4.0.\cite{PhysRevB.47.558, KRESSE199615, PhysRevB.54.11169, PhysRevB.59.1758} A kernel-based MLIP potential was constructed using a basis set derived from reference structures, which was dynamically expanded during MD simulations through an active learning approach. When the estimated Bayesian error of the predicted forces at a particular MD step exceeded a predefined threshold, the predicted forces were considered inaccurate for the current configuration. In such cases, the energy and forces of the current structure were recalculated using DFT to ensure accuracy. The DFT calculated energy and forces were used to expand the basis set of reference data and update the MLIP. The error threshold itself was adaptively updated during the MD simulation based on the average Bayesian error from the preceding ten MD steps. This approach allows the MLIP to efficiently sample the configurational phase space and identify representative structures more effectively than conventional AIMD simulations.

The interaction between ions and electrons was modelled with the projected augmented wave method\cite{PhysRevB.50.17953} with an energy cutoff of 600 eV. The convergence criterion for electronic self-consistent steps was set to 10$^{-5}$ eV. Exchange-correlation effects were treated using the BLYP functional,\cite{PhysRevA.38.3098, PhysRevB.37.785} combined with D4 dispersion corrections including Becke-Johnson damping\cite{10.1063/1.5090222}. For Brillouin zone integration, a k-point mesh with a reciprocal-space resolution of 0.25 Å$^{-1}$ was employed. All MD simulations were performed in the NVT ensemble at a target temperature of 300 K, controlled by the Langevin thermostat,\cite{PhysRevB.17.1302} with a timestep of 1.0 fs. 

\subsection{Parameterization of potential}

The energies, forces, and coordinates of structures calculated with DFT were extracted from the MD trajectories. 
%All reference energies and forces were calculated with DFT using the structures extracted from the MD trajectories.
They were used to construct a more accurate and faster MLIP with the ACE method, as implemented in the PACEMAKER package\cite{lysogorskiy2021performant, RN4, bochkarev2022efficient}. Five percent of the initial dataset was set aside as the test set, with the remaining data used for training. For the simplest systems $L_1$ and $L_5$, the ACE potential was constructed using 3000 basis functions and 8376 parameters with a maximum body order $\nu$ = 6, corresponding to seven-body interactions. For the system $S$, the ACE potential included 8000 basis functions and 20128 parameters with a maximum body order $\nu$ = 6, corresponding to seven-body interactions. For all systems containing substrate, graphene and water($S_nL_m$), the ACE potential consists of 9895 basis functions and 25430 parameters with a maximum body order $\nu$ = 5, corresponding to six-body interactions. A cutoff distance of 7.0 Å was applied across all the systems. During ACE learning, a uniform weighting scheme was applied to all the reference structures in the loss function. The initial parameterization of the potential was further refined through an active learning strategy, allowing for systematic improvement of the potential by incorporating high-error configurations encountered during MD simulations.

\subsection{Active learning } \label{AL}

\begin{figure}[ht]
\centering
\includegraphics[width=1.0\textwidth]{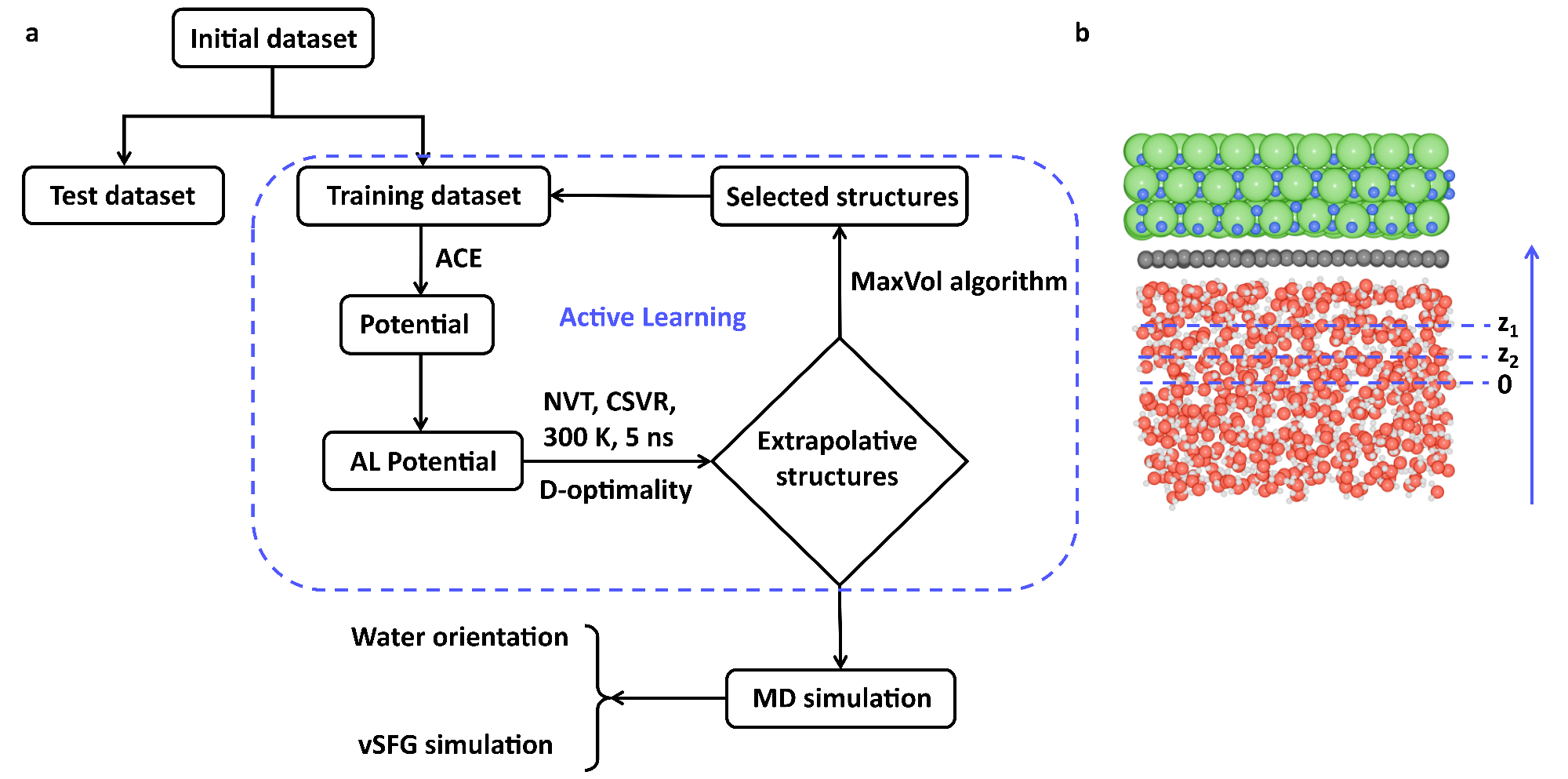}
\caption{\textbf{a}, Schematic of the MLIP training workflow: The initial interatomic potential is iteratively improved via an active learning strategy, which concludes when all structures yield extrapolation values below the specified criterion. \textbf{b}, Schematic representation of the water system for vSFG spectral simulations, where the center mass of the water system is defined as the origin, and the relative positions of $z_1$ and $z_2$ are shown.}
\label{fig7}
\end{figure}

To sample a broader range of atomic environments, we performed MD simulations in the NVT ensemble at a target temperature of 300 K, employing the canonical sampling through velocity rescaling (CSVR) thermostat\cite{10.1063/1.2408420} to control the temperature. The integration time step and damping parameters for temperature were set to 0.5 fs and 0.1 ps, respectively. The extrapolation grade $\gamma$ of each atom at every 100 simulation steps was calculated using the D-optimality criterion\cite{doi:10.1142/9789812836021_0015}. Structures with $\gamma > 5$ were selected as extrapolative structures. The MD simulations were performed with the Large-scale Atomic/Molecular Massively Parallel Simulator (LAMMPS) software\cite{LAMMPS, PLIMPTON19951} and the extrapolation grades were calculated with the ML-PACE package\cite{lysogorskiy2021performant}. To identify the most representative atomic environments, the MaxVol algorithm\cite{GOREINOV19971, doi:10.1142/9789812836021_0015} was used to construct a new active set from the structures generated in the previous steps. Structures that had multiple atoms entered the new active set were given high priority. The top candidates were selected for single-point energy calculations with DFT. The energy, forces, and coordinates of these structures were added to the previous training dataset to retrain the ACE potential. The process of MD simulation, configuration selection, single point energy calculation, and potential retraining was iteratively repeated until no structures exceeded the extrapolation grade cutoff ($\gamma \le 5$) within a 5 ns of MD steps(Fig.~\ref {fig7}a). The final ACE potential, trained with the above active learning strategy, was subsequently used for a large-system MD simulation.

\subsection{MD simulations }

MD simulations were performed using the LAMMPS package in conjunction with the ML-PACE package. Each system was first equilibrated in the NVT ensemble for 2 ns, followed by a 2 ns production run in the NVE ensemble. During the equilibration, the CSVR thermostat was used to make the system reach the target temperature of 300 K with temperature damping parameters of 0.1 ps. A fixed integration time step of 0.5 fs was used for both equilibration and production runs. Atomic coordinates were saved every 1 fs for subsequent analysis. The cell parameters and number of atoms in each system are summarized in Table S2. For the $L_5$ system, carbon atoms in the bottom two layers were kept fixed, while all other atoms were allowed to move freely. For all the systems involving the \ce{CaF2} substrate, the atoms in the bottom layer of the \ce{CaF2} slab were frozen to mimic bulk-like boundary conditions, while all the other atoms in the system were allowed to move freely during the MD simulations. 

\subsection{vSFG spectra simulations }

Since the contribution of the \ce{CaF2} substrate and graphene to vSFG spectra is negligible,\cite{10.1063/1.5016629, KIM20221187, D0CC02675A, RN62} experimentally measured vSFG signals primarily originate from interfacial water molecules. Therefore, the accuracy of simulated vSFG spectra largely depends on the precise evaluation of the dipole moments and polarizability tensors of individual water molecules. The many-body dipole moment surface (MB-$\mu$) and polarizability surface (MB-$\alpha$) have been demonstrated to reproduce the infrared and Raman spectra of liquid water accurately \cite{RN559,10.1063/1.4916629,10.1063/1.5006480}, as well as the vSFG spectra of the water–air interface \cite{RN233,RN442,RN623}. The dipole moments (“one-body + N-body”) and polarizability tensors (“one-body”) of water molecules along trajectories were computed using MB-$\mu$ and MB-$\alpha$, respectively. The total time correlation function (TCF) can be written as the sum of the autocorrelation term for the $i$th water molecule and the cross-correlation term between $i$th and $j$th water molecules, respectively. 

\begin{equation}
\left\langle\alpha_{b c}(t) \mu_{a}(0)\right\rangle=\left\langle\sum_{i} \alpha_{i, b c}(t) \mu_{i, a}(0)\right\rangle+\left\langle\sum_{i \neq j} \alpha_{i, b c}(t) \mu_{j, a}(0)\right\rangle \label{eq1}
\end{equation}

where $\alpha_{bc}$ and $\mu_a$ are the polarizability and dipole moment components of each water molecule, respectively. $\langle\cdots\rangle$ represents the thermal average. To accurately evaluate the cross-correlation contributions within the limited length of the MD trajectory, a truncated TCF was employed.\cite{RN167} A cutoff $r_t$ was introduced that only water molecule pairs ($i$ and $j$) separated by less than $r_t$ were included in the cross-correlation term. Here, the cutoff radius $r_t$ was set to 4.0 Å. 
\begin{equation}
\left \langle \alpha _{bc}\left ( t \right )\mu _{a} \left ( 0 \right )   \right \rangle =\left \langle \sum_{i}^{}\alpha _{i,bc}\left ( t \right ) \mu_{i,a}\left ( 0 \right ) \right \rangle+\left \langle \sum_{i\ne j}^{}g_{t}\left ( r_{ij};r_{t}\right )\alpha _{i,bc} \left ( t \right )\mu _{j,a} \left ( 0 \right )\right \rangle
\end{equation}

\begin{equation}
g_{t} \left ( r_{ij}\left ( t \right );r_{t}    \right ) =\begin{cases}
  1, & r_{ij}\le r_{t}  \\
  0, & r_{ij}>  r_{t}
\end{cases}
\end{equation}

For the slab model, it has two interfaces (water-air and water-graphene or \ce{CaF2}). Since the two interfaces in the simulation cell are oriented in opposite directions, and only the water–graphene or water–\ce{CaF2} interface is of interest, analysis is restricted to that specific interfacial region. A screening function $g_{sc}(z_i)$\cite{RN167, RN113} was introduced to the truncated TCF: 

\begin{align}
\left \langle \alpha _{bc}(t)\mu _{a}(0) \right \rangle 
&= \left \langle \sum_{i} \alpha _{i,bc}(t)\mu_{i,a}(0)g_{sc}^{3}(z_{i}) \right \rangle \notag \\
&\quad + \left \langle \sum_{i\ne j} g_{t}(r_{ij};r_{t}) \alpha _{i,bc}(t)\mu _{j,a}(0)g_{sc}^{3}(z_{i}) \right \rangle
%\left \langle \alpha _{bc}\left ( t \right )\mu _{a} \left ( 0 \right )   \right \rangle =\left \langle \sum_{i}^{}\alpha _{i,bc}\left ( t \right ) \mu_{i,a}\left ( 0 \right )g_{sc}^{3}\left ( z_{i}  \right )   \right \rangle+ \left \langle \sum_{i\ne j}^{}g_{t}\left ( r_{ij};r_{t}\right )\alpha _{i,bc} \left ( t \right )\mu _{j,a} \left ( 0 \right )g_{sc}^{3}\left ( z_{i}  \right )  \right \rangle
\end{align}

\begin{equation}
g_{sc}^{3}\left ( z_{i}  \right )  =\begin{cases}
  1, & z_{i}\ge  z_{1}  \\
  \cos ^{2}\left ( \frac{\pi \left ( z_{i}-z_{2}  \right ) }{2\left ( z_{1}-z_{2} \right ) }  \right ), & z_{2}\le  z_{i}< z_{1}  \\
  0, & z_{i}<   z_{2}
\end{cases}
\end{equation}

where $z_i$ is z coordinate of the $i$th water molecule. The origin point in the z direction is set to the center mass of the water system(Fig.~\ref {fig7}b). In the region $z_i \ge z_1$, the contributions from the dipole moment and polarizability are fully included in the truncated TCF. In the transition region $z_2 \le z_i < z_1$, these contributions are smoothly switched off. For other region, the contributions are set to zero.

The second-order susceptibility $\chi_{cba}^{(2)}$ can be written as a Fourier transform of the truncated TCF of the dipole moment and polarizability components,

\begin{equation}
\chi _{cba}^{\left ( 2 \right ) } \left ( \omega  \right ) =i\frac{\omega }{k_{B} T} \int_{0}^{\infty }dte^{i\omega t}\left \langle \alpha _{bc} \left ( t \right ) \mu _{a} \left ( 0 \right )  \right \rangle   
\end{equation}

where $\omega$ is the vibrational frequency, $i$ is the imaginary unit, $k_B$ is Boltzmann’s constant, $a$, $b$, and $c$ are the polarization components of the infrared IR, visible, and sum frequency beams, respectively. Since the vSFG spectra were measured with SSP polarization combination,\cite{RN91} for the simulation, we only consider the XXZ polarization combination.\cite{10.1116/6.0001401} The imaginary (Im($\chi^{(2)}$)) and real (Re($\chi^{(2)}$)) part of the vSFG spectra were computed by calculating the cosine and sine components of the squared spectrum at different surface depths. 

\subsection{Free energy calculations }\label{entropylabel}

The Gibbs free energy difference ($\Delta$G) is given by:
\begin{equation}
\Delta G = \Delta H-T\Delta S=\Delta U+p\Delta V-T\Delta S\approx \Delta U-T\Delta S
\end{equation}

where $H$ is the enthalpy, $T$ is the temperature, $S$ is the entropy, $U$ is the internal energy, $p$ is the pressure, and $V$ is the volume. Since the volume change ($\Delta$V) is negligible in our system, the Gibbs free energy difference can be approximated as the $\Delta G \approx \Delta U-T \Delta S$. The internal energy is extracted directly from LAMMPS MD simulations. To calculate the entropy of intercalated and slab water molecules, we used the two-phase thermodynamic (2PT) model\cite{RN569} implemented in DoSPT software\cite{RN570}. The internal energy and temperature used for free energy calculations correspond to the average values obtained over the first nanosecond of the 2 ns production simulation. The trajectory from the first nanosecond was also used as an input for the entropy calculations with DoSPT.

%\backmatter

%\bmhead{Supplementary information}

%If your article has accompanying supplementary files, please state so here. 

%Authors reporting data from electrophoretic gels and blots should supply the full unprocessed scans for key as part of their Supplementary information. This may be requested by the editorial team/s if it is missing.

%Please refer to Journal-level guidance for any specific requirements.

%\bmhead{Acknowledgements}
%This work was supported by the Institute for Basic Science (Grant No. IBS-R023-D1). This research was supported by the Basic Science Research Program  through the National Research Foundation of Korea (NRF) funded by the Ministry of Science and ICT, Republic of Korea(NRF-2021R1C1C1008776).

\section*{Acknowledgements}
We gratefully acknowledge Prof. Kyungwon Kwak and Dr. Jonggu Jeon for their constructive discussions. This work was supported by the Institute for Basic Science (Grant No. IBS-R023-D1). This research was supported by the Basic Science Research Program through the National Research Foundation of Korea (NRF) funded by the Ministry of Science and ICT, Republic of Korea (NRF-2021R1C1C1008776). We also acknowledge the support of the National Supercomputing Centre of Korea Institute of Science and Technology Information (KISTI) with super-computing resources, including technical support (No. KSC-2023-CRE-0468).

\section*{Declarations}

The authors declare no competing financial interest.

%\begin{appendices}

%\section{Section title of first appendix}\label{secA1}

%An appendix contains supplementary information that is not an essential part of the text itself but which may be helpful in providing a more comprehensive understanding of the research problem or it is information that is too cumbersome to be included in the body of the paper.
\section*{Data Availability}
The data that support the findings of this study are available from the corresponding author upon reasonable request.   



%\end{appendices}


\bibliography{Reference}% common bib file

\begin{thebibliography}{100}
\expandafter\ifx\csname url\endcsname\relax
  \def\url#1{\burl{#1}}\fi
\expandafter\ifx\csname urlprefix\endcsname\relax\def\urlprefix{URL }\fi
\providecommand{\bibinfo}[2]{#2}
\providecommand{\eprint}[2][]{\url{#2}}
\providecommand{\doi}[1]{\url{https://doi.org/#1}}
\bibcommenthead

\bibitem{RN592}
\bibinfo{author}{Novoselov, K.~S.} \emph{et~al.}
\newblock \bibinfo{title}{Electric field effect in atomically thin carbon films}.
\newblock \emph{\bibinfo{journal}{Science}} \textbf{\bibinfo{volume}{306}}, \bibinfo{pages}{666--669} (\bibinfo{year}{2004}).

\bibitem{RN625}
\bibinfo{author}{Avouris, P.}, \bibinfo{author}{Heinz, T.~F.} \& \bibinfo{author}{Low, T.}
\newblock \emph{\bibinfo{title}{2D Materials: Properties and Devices}}  (\bibinfo{publisher}{Cambridge University Press}, \bibinfo{address}{Cambridge}, \bibinfo{year}{2017}).

\bibitem{RN624}
\bibinfo{author}{Mas-Ballesté, R.}, \bibinfo{author}{Gómez-Navarro, C.}, \bibinfo{author}{Gómez-Herrero, J.} \& \bibinfo{author}{Zamora, F.}
\newblock \bibinfo{title}{2d materials: to graphene and beyond}.
\newblock \emph{\bibinfo{journal}{Nanoscale}} \textbf{\bibinfo{volume}{3}}, \bibinfo{pages}{20--30} (\bibinfo{year}{2011}).

\bibitem{RN596}
\bibinfo{author}{Miró, P.}, \bibinfo{author}{Audiffred, M.} \& \bibinfo{author}{Heine, T.}
\newblock \bibinfo{title}{An atlas of two-dimensional materials}.
\newblock \emph{\bibinfo{journal}{Chem. Soc. Rev.}} \textbf{\bibinfo{volume}{43}}, \bibinfo{pages}{6537--6554} (\bibinfo{year}{2014}).

\bibitem{RN648}
\bibinfo{author}{Allen, M.~J.}, \bibinfo{author}{Tung, V.~C.} \& \bibinfo{author}{Kaner, R.~B.}
\newblock \bibinfo{title}{Honeycomb carbon: A review of graphene}.
\newblock \emph{\bibinfo{journal}{Chem. Rev.}} \textbf{\bibinfo{volume}{110}}, \bibinfo{pages}{132--145} (\bibinfo{year}{2010}).

\bibitem{RN629}
\bibinfo{author}{Balandin, A.~A.}
\newblock \bibinfo{title}{Thermal properties of graphene and nanostructured carbon materials}.
\newblock \emph{\bibinfo{journal}{Nat. Mater.}} \textbf{\bibinfo{volume}{10}}, \bibinfo{pages}{569--581} (\bibinfo{year}{2011}).

\bibitem{RN630}
\bibinfo{author}{Bonaccorso, F.}, \bibinfo{author}{Sun, Z.}, \bibinfo{author}{Hasan, T.} \& \bibinfo{author}{Ferrari, A.~C.}
\newblock \bibinfo{title}{Graphene photonics and optoelectronics}.
\newblock \emph{\bibinfo{journal}{Nat. Photonics}} \textbf{\bibinfo{volume}{4}}, \bibinfo{pages}{611--622} (\bibinfo{year}{2010}).

\bibitem{RN627}
\bibinfo{author}{Castro~Neto, A.~H.}, \bibinfo{author}{Guinea, F.}, \bibinfo{author}{Peres, N. M.~R.}, \bibinfo{author}{Novoselov, K.~S.} \& \bibinfo{author}{Geim, A.~K.}
\newblock \bibinfo{title}{The electronic properties of graphene}.
\newblock \emph{\bibinfo{journal}{Rev. Mod. Phys.}} \textbf{\bibinfo{volume}{81}}, \bibinfo{pages}{109--162} (\bibinfo{year}{2009}).

\bibitem{RN628}
\bibinfo{author}{Sun, Y.~W.} \emph{et~al.}
\newblock \bibinfo{title}{Mechanical properties of graphene}.
\newblock \emph{\bibinfo{journal}{Appl. Phys. Rev.}} \textbf{\bibinfo{volume}{8}}, \bibinfo{pages}{021310} (\bibinfo{year}{2021}).

\bibitem{XU201873}
\bibinfo{author}{Xu, Z.}
\newblock \bibinfo{title}{ in \textit{Fundamental properties of graphene}} (eds \bibinfo{editor}{Zhu, H.}, \bibinfo{editor}{Xu, Z.}, \bibinfo{editor}{Xie, D.} \& \bibinfo{editor}{Fang, Y.}) \emph{\bibinfo{booktitle}{Graphene}} \bibinfo{pages}{73--102} (\bibinfo{publisher}{Academic Press}, \bibinfo{address}{Cambridge, MA}, \bibinfo{year}{2018}).

\bibitem{RN631}
\bibinfo{author}{Zhao, G.} \emph{et~al.}
\newblock \bibinfo{title}{The physics and chemistry of graphene-on-surfaces}.
\newblock \emph{\bibinfo{journal}{Chem. Soc. Rev.}} \textbf{\bibinfo{volume}{46}}, \bibinfo{pages}{4417--4449} (\bibinfo{year}{2017}).

\bibitem{RN646}
\bibinfo{author}{Chen, K.}, \bibinfo{author}{Song, S.}, \bibinfo{author}{Liu, F.} \& \bibinfo{author}{Xue, D.}
\newblock \bibinfo{title}{Structural design of graphene for use in electrochemical energy storage devices}.
\newblock \emph{\bibinfo{journal}{Chem. Soc. Rev.}} \textbf{\bibinfo{volume}{44}}, \bibinfo{pages}{6230--6257} (\bibinfo{year}{2015}).

\bibitem{RN643}
\bibinfo{author}{Kang, J.} \emph{et~al.}
\newblock \bibinfo{title}{Graphene membrane for water-related environmental application: A comprehensive review and perspectives}.
\newblock \emph{\bibinfo{journal}{ACS Environ. Au}} \textbf{\bibinfo{volume}{5}}, \bibinfo{pages}{35--60} (\bibinfo{year}{2025}).

\bibitem{RN647}
\bibinfo{author}{Liu, M.}, \bibinfo{author}{Zhang, R.} \& \bibinfo{author}{Chen, W.}
\newblock \bibinfo{title}{Graphene-supported nanoelectrocatalysts for fuel cells: Synthesis, properties, and applications}.
\newblock \emph{\bibinfo{journal}{Chem. Rev.}} \textbf{\bibinfo{volume}{114}}, \bibinfo{pages}{5117--5160} (\bibinfo{year}{2014}).

\bibitem{RN644}
\bibinfo{author}{Xu, Y.} \& \bibinfo{author}{Liu, J.}
\newblock \bibinfo{title}{Graphene as transparent electrodes: Fabrication and new emerging applications}.
\newblock \emph{\bibinfo{journal}{Small}} \textbf{\bibinfo{volume}{12}}, \bibinfo{pages}{1400--1419} (\bibinfo{year}{2016}).

\bibitem{RN645}
\bibinfo{author}{Zhu, J.}, \bibinfo{author}{Yang, D.}, \bibinfo{author}{Yin, Z.}, \bibinfo{author}{Yan, Q.} \& \bibinfo{author}{Zhang, H.}
\newblock \bibinfo{title}{Graphene and graphene-based materials for energy storage applications}.
\newblock \emph{\bibinfo{journal}{Small}} \textbf{\bibinfo{volume}{10}}, \bibinfo{pages}{3480--3498} (\bibinfo{year}{2014}).

\bibitem{RN641}
\bibinfo{author}{Angizi, S.}, \bibinfo{author}{Hong, L.}, \bibinfo{author}{Huang, X.}, \bibinfo{author}{Selvaganapathy, P.~R.} \& \bibinfo{author}{Kruse, P.}
\newblock \bibinfo{title}{Graphene versus concentrated aqueous electrolytes: the role of the electrochemical double layer in determining the screening length of an electrolyte}.
\newblock \emph{\bibinfo{journal}{npj 2D Mater. Appl.}} \textbf{\bibinfo{volume}{7}}, \bibinfo{pages}{67} (\bibinfo{year}{2023}).

\bibitem{RN642}
\bibinfo{author}{Elliott, J.~D.}, \bibinfo{author}{Papaderakis, A.~A.}, \bibinfo{author}{Dryfe, R. A.~W.} \& \bibinfo{author}{Carbone, P.}
\newblock \bibinfo{title}{The electrochemical double layer at the graphene/aqueous electrolyte interface: what we can learn from simulations, experiments, and theory}.
\newblock \emph{\bibinfo{journal}{J. Mater. Chem. C}} \textbf{\bibinfo{volume}{10}}, \bibinfo{pages}{15225--15262} (\bibinfo{year}{2022}).

\bibitem{RN53}
\bibinfo{author}{Garcia, R.}
\newblock \bibinfo{title}{Interfacial liquid water on graphite, graphene, and 2d materials}.
\newblock \emph{\bibinfo{journal}{ACS Nano}} \textbf{\bibinfo{volume}{17}}, \bibinfo{pages}{51--69} (\bibinfo{year}{2023}).

\bibitem{B808149M}
\bibinfo{author}{Devanathan, R.}
\newblock \bibinfo{title}{Recent developments in proton exchange membranes for fuel cells}.
\newblock \emph{\bibinfo{journal}{Energy Environ. Sci.}} \textbf{\bibinfo{volume}{1}}, \bibinfo{pages}{101--119} (\bibinfo{year}{2008}).

\bibitem{moilanen2008water}
\bibinfo{author}{Moilanen, D.~E.}, \bibinfo{author}{Spry, D.} \& \bibinfo{author}{Fayer, M.}
\newblock \bibinfo{title}{Water dynamics and proton transfer in nafion fuel cell membranes}.
\newblock \emph{\bibinfo{journal}{Langmuir}} \textbf{\bibinfo{volume}{24}}, \bibinfo{pages}{3690--3698} (\bibinfo{year}{2008}).

\bibitem{xu2019transparent}
\bibinfo{author}{Xu, J.} \emph{et~al.}
\newblock \bibinfo{title}{Transparent proton transport through a two-dimensional nanomesh material}.
\newblock \emph{\bibinfo{journal}{Nat. Commun.}} \textbf{\bibinfo{volume}{10}}, \bibinfo{pages}{3971} (\bibinfo{year}{2019}).

\bibitem{sachar2021hydrogen}
\bibinfo{author}{Sachar, H.~S.}, \bibinfo{author}{Chava, B.~S.}, \bibinfo{author}{Pial, T.~H.} \& \bibinfo{author}{Das, S.}
\newblock \bibinfo{title}{Hydrogen bonding and its effect on the orientational dynamics of water molecules inside polyelectrolyte brush-induced soft and active nanoconfinement}.
\newblock \emph{\bibinfo{journal}{Macromolecules}} \textbf{\bibinfo{volume}{54}}, \bibinfo{pages}{2011--2021} (\bibinfo{year}{2021}).

\bibitem{joly2016strong}
\bibinfo{author}{Joly, L.}, \bibinfo{author}{Tocci, G.}, \bibinfo{author}{Merabia, S.} \& \bibinfo{author}{Michaelides, A.}
\newblock \bibinfo{title}{Strong coupling between nanofluidic transport and interfacial chemistry: How defect reactivity controls liquid--solid friction through hydrogen bonding}.
\newblock \emph{\bibinfo{journal}{J. Phys. Chem. Lett.}} \textbf{\bibinfo{volume}{7}}, \bibinfo{pages}{1381--1386} (\bibinfo{year}{2016}).

\bibitem{RN609}
\bibinfo{author}{Eral, H.~B.}, \bibinfo{author}{’t Mannetje, D. J. C.~M.} \& \bibinfo{author}{Oh, J.~M.}
\newblock \bibinfo{title}{Contact angle hysteresis: a review of fundamentals and applications}.
\newblock \emph{\bibinfo{journal}{Colloid Polym. Sci.}} \textbf{\bibinfo{volume}{291}}, \bibinfo{pages}{247--260} (\bibinfo{year}{2013}).

\bibitem{RN607}
\bibinfo{author}{Good, R.~J.}
\newblock \bibinfo{title}{Contact angle, wetting, and adhesion: a critical review}.
\newblock \emph{\bibinfo{journal}{J. Adhes. Sci. Technol.}} \textbf{\bibinfo{volume}{6}}, \bibinfo{pages}{1269--1302} (\bibinfo{year}{1992}).

\bibitem{RN608}
\bibinfo{author}{Song, J.-W.} \& \bibinfo{author}{Fan, L.-W.}
\newblock \bibinfo{title}{Temperature dependence of the contact angle of water: A review of research progress, theoretical understanding, and implications for boiling heat transfer}.
\newblock \emph{\bibinfo{journal}{Adv. Colloid Interface Sci.}} \textbf{\bibinfo{volume}{288}}, \bibinfo{pages}{102339} (\bibinfo{year}{2021}).

\bibitem{RN586}
\bibinfo{author}{Belyaeva, L.~A.}, \bibinfo{author}{Tang, C.}, \bibinfo{author}{Juurlink, L.} \& \bibinfo{author}{Schneider, G.~F.}
\newblock \bibinfo{title}{Macroscopic and microscopic wettability of graphene}.
\newblock \emph{\bibinfo{journal}{Langmuir}} \textbf{\bibinfo{volume}{37}}, \bibinfo{pages}{4049--4055} (\bibinfo{year}{2021}).

\bibitem{RN499}
\bibinfo{author}{Driskill, J.}, \bibinfo{author}{Vanzo, D.}, \bibinfo{author}{Bratko, D.} \& \bibinfo{author}{Luzar, A.}
\newblock \bibinfo{title}{Wetting transparency of graphene in water}.
\newblock \emph{\bibinfo{journal}{J. Chem. Phys.}} \textbf{\bibinfo{volume}{141}}, \bibinfo{pages}{18C517} (\bibinfo{year}{2014}).

\bibitem{RN496}
\bibinfo{author}{Rafiee, J.} \emph{et~al.}
\newblock \bibinfo{title}{Wetting transparency of graphene}.
\newblock \emph{\bibinfo{journal}{Nat. Mater.}} \textbf{\bibinfo{volume}{11}}, \bibinfo{pages}{217--222} (\bibinfo{year}{2012}).

\bibitem{RN616}
\bibinfo{author}{Ashraf, A.} \emph{et~al.}
\newblock \bibinfo{title}{Doping-induced tunable wettability and adhesion of graphene}.
\newblock \emph{\bibinfo{journal}{Nano Lett.}} \textbf{\bibinfo{volume}{16}}, \bibinfo{pages}{4708--4712} (\bibinfo{year}{2016}).

\bibitem{RN615}
\bibinfo{author}{Hong, G.} \emph{et~al.}
\newblock \bibinfo{title}{On the mechanism of hydrophilicity of graphene}.
\newblock \emph{\bibinfo{journal}{Nano Lett.}} \textbf{\bibinfo{volume}{16}}, \bibinfo{pages}{4447--4453} (\bibinfo{year}{2016}).

\bibitem{RN612}
\bibinfo{author}{Kozbial, A.}, \bibinfo{author}{Trouba, C.}, \bibinfo{author}{Liu, H.} \& \bibinfo{author}{Li, L.}
\newblock \bibinfo{title}{Characterization of the intrinsic water wettability of graphite using contact angle measurements: Effect of defects on static and dynamic contact angles}.
\newblock \emph{\bibinfo{journal}{Langmuir}} \textbf{\bibinfo{volume}{33}}, \bibinfo{pages}{959--967} (\bibinfo{year}{2017}).

\bibitem{RN613}
\bibinfo{author}{Li, X.}, \bibinfo{author}{Li, L.}, \bibinfo{author}{Wang, Y.}, \bibinfo{author}{Li, H.} \& \bibinfo{author}{Bian, X.}
\newblock \bibinfo{title}{Wetting and interfacial properties of water on the defective graphene}.
\newblock \emph{\bibinfo{journal}{J. Phys. Chem. C}} \textbf{\bibinfo{volume}{117}}, \bibinfo{pages}{14106--14112} (\bibinfo{year}{2013}).

\bibitem{RN91}
\bibinfo{author}{Kim, D.} \emph{et~al.}
\newblock \bibinfo{title}{Wettability of graphene and interfacial water structure}.
\newblock \emph{\bibinfo{journal}{Chem}} \textbf{\bibinfo{volume}{7}}, \bibinfo{pages}{1602--1614} (\bibinfo{year}{2021}).

\bibitem{doi:10.1021/acs.jpcc.5b10492}
\bibinfo{author}{Aria, A.~I.} \emph{et~al.}
\newblock \bibinfo{title}{Time evolution of the wettability of supported graphene under ambient air exposure}.
\newblock \emph{\bibinfo{journal}{J. Phys. Chem. C}} \textbf{\bibinfo{volume}{120}}, \bibinfo{pages}{2215--2224} (\bibinfo{year}{2016}).

\bibitem{RN599}
\bibinfo{author}{Kozbial, A.} \emph{et~al.}
\newblock \bibinfo{title}{Study on the surface energy of graphene by contact angle measurements}.
\newblock \emph{\bibinfo{journal}{Langmuir}} \textbf{\bibinfo{volume}{30}}, \bibinfo{pages}{8598--8606} (\bibinfo{year}{2014}).

\bibitem{RN614}
\bibinfo{author}{Li, Z.} \emph{et~al.}
\newblock \bibinfo{title}{Effect of airborne contaminants on the wettability of supported graphene and graphite}.
\newblock \emph{\bibinfo{journal}{Nat. Mater.}} \textbf{\bibinfo{volume}{12}}, \bibinfo{pages}{925--931} (\bibinfo{year}{2013}).

\bibitem{RN619}
\bibinfo{author}{Carlson, S.~R.}, \bibinfo{author}{Schullian, O.}, \bibinfo{author}{Becker, M.~R.} \& \bibinfo{author}{Netz, R.~R.}
\newblock \bibinfo{title}{Modeling water interactions with graphene and graphite via force fields consistent with experimental contact angles}.
\newblock \emph{\bibinfo{journal}{J. Phys. Chem. Lett.}} \textbf{\bibinfo{volume}{15}}, \bibinfo{pages}{6325--6333} (\bibinfo{year}{2024}).

\bibitem{RN618}
\bibinfo{author}{Prydatko, A.~V.}, \bibinfo{author}{Belyaeva, L.~A.}, \bibinfo{author}{Jiang, L.}, \bibinfo{author}{Lima, L. M.~C.} \& \bibinfo{author}{Schneider, G.~F.}
\newblock \bibinfo{title}{Contact angle measurement of free-standing square-millimeter single-layer graphene}.
\newblock \emph{\bibinfo{journal}{Nat. Commun.}} \textbf{\bibinfo{volume}{9}}, \bibinfo{pages}{4185} (\bibinfo{year}{2018}).

\bibitem{RN600}
\bibinfo{author}{Taherian, F.}, \bibinfo{author}{Marcon, V.}, \bibinfo{author}{van~der Vegt, N. F.~A.} \& \bibinfo{author}{Leroy, F.}
\newblock \bibinfo{title}{What is the contact angle of water on graphene?}
\newblock \emph{\bibinfo{journal}{Langmuir}} \textbf{\bibinfo{volume}{29}}, \bibinfo{pages}{1457--1465} (\bibinfo{year}{2013}).

\bibitem{BELYAEVA2020100482}
\bibinfo{author}{Belyaeva, L.~A.} \& \bibinfo{author}{Schneider, G.~F.}
\newblock \bibinfo{title}{Wettability of graphene}.
\newblock \emph{\bibinfo{journal}{Surf. Sci. Rep.}} \textbf{\bibinfo{volume}{75}}, \bibinfo{pages}{100482} (\bibinfo{year}{2020}).

\bibitem{RN621}
\bibinfo{author}{Kawaguchi, D.} \& \bibinfo{author}{Tanaka, K.}
\newblock \bibinfo{title}{ in \textit{Sum frequency generation (sfg)}} (eds \bibinfo{editor}{Maeda, M.}, \bibinfo{editor}{Takahara, A.}, \bibinfo{editor}{Kitano, H.}, \bibinfo{editor}{Yamaoka, T.} \& \bibinfo{editor}{Miura, Y.}) \emph{\bibinfo{booktitle}{Molecular Soft-Interface Science: Principles, Molecular Design, Characterization and Application}} \bibinfo{pages}{87--99} (\bibinfo{publisher}{Springer Japan}, \bibinfo{address}{Tokyo}, \bibinfo{year}{2019}).

\bibitem{RN620}
\bibinfo{author}{Nihonyanagi, S.} \& \bibinfo{author}{Tahara, T.}
\newblock \bibinfo{title}{ in \textit{Vibrational sum frequency generation spectroscopy}} (ed.\bibinfo{editor}{The Surface Science Society~of, J.}) \emph{\bibinfo{booktitle}{Compendium of Surface and Interface Analysis}} \bibinfo{pages}{801--807} (\bibinfo{publisher}{Springer Singapore}, \bibinfo{address}{Singapore}, \bibinfo{year}{2018}).

\bibitem{RN26}
\bibinfo{author}{Pickering, J.~D.} \emph{et~al.}
\newblock \bibinfo{title}{Tutorials in vibrational sum frequency generation spectroscopy. ii. designing a broadband vibrational sum frequency generation spectrometer}.
\newblock \emph{\bibinfo{journal}{Biointerphases}} \textbf{\bibinfo{volume}{17}}, \bibinfo{pages}{011202} (\bibinfo{year}{2022}).

\bibitem{RN15}
\bibinfo{author}{Pickering, J.~D.} \emph{et~al.}
\newblock \bibinfo{title}{Tutorials in vibrational sum frequency generation spectroscopy. i. the foundations}.
\newblock \emph{\bibinfo{journal}{Biointerphases}} \textbf{\bibinfo{volume}{17}}, \bibinfo{pages}{011201} (\bibinfo{year}{2022}).

\bibitem{RN622}
\bibinfo{author}{Pramhaas, V.} \& \bibinfo{author}{Rupprechter, G.}
\newblock \bibinfo{title}{ in \textit{Sum frequency generation (sfg) spectroscopy}} (eds \bibinfo{editor}{Wachs, I.~E.} \& \bibinfo{editor}{Bañares, M.~A.}) \emph{\bibinfo{booktitle}{Springer Handbook of Advanced Catalyst Characterization}} \bibinfo{pages}{213--233} (\bibinfo{publisher}{Springer International Publishing}, \bibinfo{address}{Cham}, \bibinfo{year}{2023}).

\bibitem{RN606}
\bibinfo{author}{Gan, W.}, \bibinfo{author}{Wu, D.}, \bibinfo{author}{Zhang, Z.}, \bibinfo{author}{Feng, R.-r.} \& \bibinfo{author}{Wang, H.-f.}
\newblock \bibinfo{title}{Polarization and experimental configuration analyses of sum frequency generation vibrational spectra, structure, and orientational motion of the air/water interface}.
\newblock \emph{\bibinfo{journal}{J. Chem. Phys.}} \textbf{\bibinfo{volume}{124}}, \bibinfo{pages}{114705} (\bibinfo{year}{2006}).

\bibitem{RN103}
\bibinfo{author}{Hong, Y.}, \bibinfo{author}{He, J.}, \bibinfo{author}{Zhang, C.} \& \bibinfo{author}{Wang, X.}
\newblock \bibinfo{title}{Probing the structure of water at the interface with graphene oxide using sum frequency generation vibrational spectroscopy}.
\newblock \emph{\bibinfo{journal}{J. Phys. Chem. C}} \textbf{\bibinfo{volume}{126}}, \bibinfo{pages}{1471--1480} (\bibinfo{year}{2022}).

\bibitem{RN605}
\bibinfo{author}{Nihonyanagi, S.}, \bibinfo{author}{Mondal, J.~A.}, \bibinfo{author}{Yamaguchi, S.} \& \bibinfo{author}{Tahara, T.}
\newblock \bibinfo{title}{Structure and dynamics of interfacial water studied by heterodyne-detected vibrational sum-frequency generation}.
\newblock \emph{\bibinfo{journal}{Annu. Rev. Phys. Chem.}} \textbf{\bibinfo{volume}{64}}, \bibinfo{pages}{579--603} (\bibinfo{year}{2013}).

\bibitem{RN604}
\bibinfo{author}{Nihonyanagi, S.}, \bibinfo{author}{Yamaguchi, S.} \& \bibinfo{author}{Tahara, T.}
\newblock \bibinfo{title}{Ultrafast dynamics at water interfaces studied by vibrational sum frequency generation spectroscopy}.
\newblock \emph{\bibinfo{journal}{Chem. Rev.}} \textbf{\bibinfo{volume}{117}}, \bibinfo{pages}{10665--10693} (\bibinfo{year}{2017}).

\bibitem{RN603}
\bibinfo{author}{Perakis, F.} \emph{et~al.}
\newblock \bibinfo{title}{Vibrational spectroscopy and dynamics of water}.
\newblock \emph{\bibinfo{journal}{Chem. Rev.}} \textbf{\bibinfo{volume}{116}}, \bibinfo{pages}{7590--7607} (\bibinfo{year}{2016}).

\bibitem{RN602}
\bibinfo{author}{Shen, Y.~R.} \& \bibinfo{author}{Ostroverkhov, V.}
\newblock \bibinfo{title}{Sum-frequency vibrational spectroscopy on water interfaces: Polar orientation of water molecules at interfaces}.
\newblock \emph{\bibinfo{journal}{Chem. Rev.}} \textbf{\bibinfo{volume}{106}}, \bibinfo{pages}{1140--1154} (\bibinfo{year}{2006}).

\bibitem{RN107}
\bibinfo{author}{Singla, S.} \emph{et~al.}
\newblock \bibinfo{title}{Insight on structure of water and ice next to graphene using surface-sensitive spectroscopy}.
\newblock \emph{\bibinfo{journal}{ACS Nano}} \textbf{\bibinfo{volume}{11}}, \bibinfo{pages}{4899--4906} (\bibinfo{year}{2017}).

\bibitem{RN493}
\bibinfo{author}{Wang, Y.} \emph{et~al.}
\newblock \bibinfo{title}{Heterodyne-detected sum-frequency generation vibrational spectroscopy reveals aqueous molecular structure at the suspended graphene/water interface}.
\newblock \emph{\bibinfo{journal}{Angew. Chem.}} \textbf{\bibinfo{volume}{63}}, \bibinfo{pages}{e202319503} (\bibinfo{year}{2024}).

\bibitem{RN669}
\bibinfo{author}{Xu, Y.}, \bibinfo{author}{Ma, Y.-B.}, \bibinfo{author}{Gu, F.}, \bibinfo{author}{Yang, S.-S.} \& \bibinfo{author}{Tian, C.-S.}
\newblock \bibinfo{title}{Structure evolution at the gate-tunable suspended graphene–water interface}.
\newblock \emph{\bibinfo{journal}{Nature}} \textbf{\bibinfo{volume}{621}}, \bibinfo{pages}{506--510} (\bibinfo{year}{2023}).

\bibitem{RN97}
\bibinfo{author}{Dreier, L.~B.} \emph{et~al.}
\newblock \bibinfo{title}{Surface-specific spectroscopy of water at a potentiostatically controlled supported graphene monolayer}.
\newblock \emph{\bibinfo{journal}{J. Phys. Chem. C}} \textbf{\bibinfo{volume}{123}}, \bibinfo{pages}{24031--24038} (\bibinfo{year}{2019}).

\bibitem{RN551}
\bibinfo{author}{Khatib, R.} \emph{et~al.}
\newblock \bibinfo{title}{Water orientation and hydrogen-bond structure at the fluorite/water interface}.
\newblock \emph{\bibinfo{journal}{Sci. Rep.}} \textbf{\bibinfo{volume}{6}}, \bibinfo{pages}{24287} (\bibinfo{year}{2016}).

\bibitem{RN525}
\bibinfo{author}{Lesnicki, D.}, \bibinfo{author}{Zhang, Z.}, \bibinfo{author}{Bonn, M.}, \bibinfo{author}{Sulpizi, M.} \& \bibinfo{author}{Backus, E. H.~G.}
\newblock \bibinfo{title}{Surface charges at the \ce{CaF2}/water interface allow very fast intermolecular vibrational-energy transfer}.
\newblock \emph{\bibinfo{journal}{Angew. Chem.}} \textbf{\bibinfo{volume}{59}}, \bibinfo{pages}{13116--13121} (\bibinfo{year}{2020}).

\bibitem{RN503}
\bibinfo{author}{Yang, F.} \emph{et~al.}
\newblock \bibinfo{title}{Wetting transparency of single-layer graphene on liquid substrates}.
\newblock \emph{\bibinfo{journal}{Adv. Mater.}} \textbf{\bibinfo{volume}{36}}, \bibinfo{pages}{2403820} (\bibinfo{year}{2024}).

\bibitem{RN504}
\bibinfo{author}{Zhang, J.} \emph{et~al.}
\newblock \bibinfo{title}{Intrinsic wettability in pristine graphene}.
\newblock \emph{\bibinfo{journal}{Adv. Mater.}} \textbf{\bibinfo{volume}{34}}, \bibinfo{pages}{2103620} (\bibinfo{year}{2022}).

\bibitem{RN598}
\bibinfo{author}{Du, F.}, \bibinfo{author}{Huang, J.}, \bibinfo{author}{Duan, H.}, \bibinfo{author}{Xiong, C.} \& \bibinfo{author}{Wang, J.}
\newblock \bibinfo{title}{Wetting transparency of supported graphene is regulated by polarities of liquids and substrates}.
\newblock \emph{\bibinfo{journal}{Appl. Surf. Sci.}} \textbf{\bibinfo{volume}{454}}, \bibinfo{pages}{249--255} (\bibinfo{year}{2018}).

\bibitem{RN500}
\bibinfo{author}{Kim, D.}, \bibinfo{author}{Pugno, N.~M.}, \bibinfo{author}{Buehler, M.~J.} \& \bibinfo{author}{Ryu, S.}
\newblock \bibinfo{title}{Solving the controversy on the wetting transparency of graphene}.
\newblock \emph{\bibinfo{journal}{Sci. Rep.}} \textbf{\bibinfo{volume}{5}}, \bibinfo{pages}{15526} (\bibinfo{year}{2015}).

\bibitem{RN611}
\bibinfo{author}{Shih, C.-J.} \emph{et~al.}
\newblock \bibinfo{title}{Breakdown in the wetting transparency of graphene}.
\newblock \emph{\bibinfo{journal}{Phys. Rev. Lett.}} \textbf{\bibinfo{volume}{109}}, \bibinfo{pages}{176101} (\bibinfo{year}{2012}).

\bibitem{RN589}
\bibinfo{author}{Raj, R.}, \bibinfo{author}{Maroo, S.~C.} \& \bibinfo{author}{Wang, E.~N.}
\newblock \bibinfo{title}{Wettability of graphene}.
\newblock \emph{\bibinfo{journal}{Nano Lett.}} \textbf{\bibinfo{volume}{13}}, \bibinfo{pages}{1509--1515} (\bibinfo{year}{2013}).

\bibitem{RN501}
\bibinfo{author}{Shih, C.-J.}, \bibinfo{author}{Strano, M.~S.} \& \bibinfo{author}{Blankschtein, D.}
\newblock \bibinfo{title}{Wetting translucency of graphene}.
\newblock \emph{\bibinfo{journal}{Nat. Mater.}} \textbf{\bibinfo{volume}{12}}, \bibinfo{pages}{866--869} (\bibinfo{year}{2013}).

\bibitem{RN601}
\bibinfo{author}{Shin, Y.~J.} \emph{et~al.}
\newblock \bibinfo{title}{Surface-energy engineering of graphene}.
\newblock \emph{\bibinfo{journal}{Langmuir}} \textbf{\bibinfo{volume}{26}}, \bibinfo{pages}{3798--3802} (\bibinfo{year}{2010}).

\bibitem{D0NR08843A}
\bibinfo{author}{Gaire, B.}, \bibinfo{author}{Singla, S.} \& \bibinfo{author}{Dhinojwala, A.}
\newblock \bibinfo{title}{Screening of hydrogen bonding interactions by a single layer graphene}.
\newblock \emph{\bibinfo{journal}{Nanoscale}} \textbf{\bibinfo{volume}{13}}, \bibinfo{pages}{8098--8106} (\bibinfo{year}{2021}).

\bibitem{RN52}
\bibinfo{author}{Ohto, T.}, \bibinfo{author}{Tada, H.} \& \bibinfo{author}{Nagata, Y.}
\newblock \bibinfo{title}{Structure and dynamics of water at water–graphene and water–hexagonal boron-nitride sheet interfaces revealed by ab initio sum-frequency generation spectroscopy}.
\newblock \emph{\bibinfo{journal}{Phys. Chem. Chem. Phys.}} \textbf{\bibinfo{volume}{20}}, \bibinfo{pages}{12979--12985} (\bibinfo{year}{2018}).

\bibitem{RN515}
\bibinfo{author}{Bollmann, T. R.~J.}, \bibinfo{author}{Antipina, L.~Y.}, \bibinfo{author}{Temmen, M.}, \bibinfo{author}{Reichling, M.} \& \bibinfo{author}{Sorokin, P.~B.}
\newblock \bibinfo{title}{Hole-doping of mechanically exfoliated graphene by confined hydration layers}.
\newblock \emph{\bibinfo{journal}{Nano Res.}} \textbf{\bibinfo{volume}{8}}, \bibinfo{pages}{3020--3026} (\bibinfo{year}{2015}).

\bibitem{RN74}
\bibinfo{author}{Temmen, M.}, \bibinfo{author}{Ochedowski, O.}, \bibinfo{author}{Schleberger, M.}, \bibinfo{author}{Reichling, M.} \& \bibinfo{author}{Bollmann, T. R.~J.}
\newblock \bibinfo{title}{Hydration layers trapped between graphene and a hydrophilic substrate}.
\newblock \emph{\bibinfo{journal}{New J. Phys.}} \textbf{\bibinfo{volume}{16}}, \bibinfo{pages}{053039} (\bibinfo{year}{2014}).

\bibitem{RN528}
\bibinfo{author}{Zhao, W.-H.}, \bibinfo{author}{Bai, J.}, \bibinfo{author}{Yuan, L.-F.}, \bibinfo{author}{Yang, J.} \& \bibinfo{author}{Zeng, X.~C.}
\newblock \bibinfo{title}{Ferroelectric hexagonal and rhombic monolayer ice phases}.
\newblock \emph{\bibinfo{journal}{Chem. Sci.}} \textbf{\bibinfo{volume}{5}}, \bibinfo{pages}{1757--1764} (\bibinfo{year}{2014}).

\bibitem{RN1}
\bibinfo{author}{Kaliannan, N.~K.} \emph{et~al.}
\newblock \bibinfo{title}{Impact of intermolecular vibrational coupling effects on the sum-frequency generation spectra of the water/air interface}.
\newblock \emph{\bibinfo{journal}{Mol. Phys.}} \textbf{\bibinfo{volume}{118}}, \bibinfo{pages}{1620358} (\bibinfo{year}{2020}).

\bibitem{RN157}
\bibinfo{author}{Liang, C.}, \bibinfo{author}{Jeon, J.} \& \bibinfo{author}{Cho, M.}
\newblock \bibinfo{title}{Ab initio modeling of the vibrational sum-frequency generation spectrum of interfacial water}.
\newblock \emph{\bibinfo{journal}{J. Phys. Chem. Lett.}} \textbf{\bibinfo{volume}{10}}, \bibinfo{pages}{1153} (\bibinfo{year}{2019}).

\bibitem{RN43}
\bibinfo{author}{Li, X.} \emph{et~al.}
\newblock \bibinfo{title}{Influence of water on the electronic structure of metal-supported graphene: Insights from van der waals density functional theory}.
\newblock \emph{\bibinfo{journal}{Phys. Rev. B}} \textbf{\bibinfo{volume}{85}}, \bibinfo{pages}{085425} (\bibinfo{year}{2012}).

\bibitem{RN80}
\bibinfo{author}{Inagaki, T.}, \bibinfo{author}{Hatanaka, M.} \& \bibinfo{author}{Saito, S.}
\newblock \bibinfo{title}{Anisotropic and finite effects on intermolecular vibration and relaxation dynamics: Low-frequency raman spectroscopy of water film and droplet on graphene by molecular dynamics simulations}.
\newblock \emph{\bibinfo{journal}{J. Phys. Chem. B}} \textbf{\bibinfo{volume}{127}}, \bibinfo{pages}{5869--5880} (\bibinfo{year}{2023}).

\bibitem{RN35}
\bibinfo{author}{Akaishi, A.}, \bibinfo{author}{Yonemaru, T.} \& \bibinfo{author}{Nakamura, J.}
\newblock \bibinfo{title}{Formation of water layers on graphene surfaces}.
\newblock \emph{\bibinfo{journal}{ACS Omega}} \textbf{\bibinfo{volume}{2}}, \bibinfo{pages}{2184--2190} (\bibinfo{year}{2017}).

\bibitem{RN623}
\bibinfo{author}{Rashmi, R.} \emph{et~al.}
\newblock \bibinfo{title}{Revealing the water structure at neutral and charged graphene/water interfaces through quantum simulations of sum frequency generation spectra}.
\newblock \emph{\bibinfo{journal}{ACS Nano}} \textbf{\bibinfo{volume}{19}}, \bibinfo{pages}{4876--4886} (\bibinfo{year}{2025}).

\bibitem{RN591}
\bibinfo{author}{Montenegro, A.} \emph{et~al.}
\newblock \bibinfo{title}{Asymmetric response of interfacial water to applied electric fields}.
\newblock \emph{\bibinfo{journal}{Nature}} \textbf{\bibinfo{volume}{594}}, \bibinfo{pages}{62--65} (\bibinfo{year}{2021}).

\bibitem{munz2015thickness}
\bibinfo{author}{Munz, M.}, \bibinfo{author}{Giusca, C.~E.}, \bibinfo{author}{Myers-Ward, R.~L.}, \bibinfo{author}{Gaskill, D.~K.} \& \bibinfo{author}{Kazakova, O.}
\newblock \bibinfo{title}{Thickness-dependent hydrophobicity of epitaxial graphene}.
\newblock \emph{\bibinfo{journal}{ACS Nano}} \textbf{\bibinfo{volume}{9}}, \bibinfo{pages}{8401--8411} (\bibinfo{year}{2015}).

\bibitem{D3FD00107E}
\bibinfo{author}{Wang, Y.}, \bibinfo{author}{Nagata, Y.} \& \bibinfo{author}{Bonn, M.}
\newblock \bibinfo{title}{Substrate effect on charging of electrified graphene/water interfaces}.
\newblock \emph{\bibinfo{journal}{Faraday Discuss.}} \textbf{\bibinfo{volume}{249}}, \bibinfo{pages}{303--316} (\bibinfo{year}{2024}).

\bibitem{RN62}
\bibinfo{author}{Wang, Y.} \emph{et~al.}
\newblock \bibinfo{title}{Direct probe of electrochemical pseudocapacitive ph jump at a graphene electrode**}.
\newblock \emph{\bibinfo{journal}{Angew. Chem.}} \textbf{\bibinfo{volume}{62}}, \bibinfo{pages}{e202216604} (\bibinfo{year}{2023}).

\bibitem{RN60}
\bibinfo{author}{Wang, Y.} \emph{et~al.}
\newblock \bibinfo{title}{Chemistry governs water organization at a graphene electrode}.
\newblock \emph{\bibinfo{journal}{Nature}} \textbf{\bibinfo{volume}{615}}, \bibinfo{pages}{E1--E2} (\bibinfo{year}{2023}).

\bibitem{RN662}
\bibinfo{author}{Wang, Y.} \emph{et~al.}
\newblock \bibinfo{title}{Interfaces govern the structure of angstrom-scale confined water solutions}.
\newblock \emph{\bibinfo{journal}{Nat. Commun.}} \textbf{\bibinfo{volume}{16}}, \bibinfo{pages}{7288} (\bibinfo{year}{2025}).

\bibitem{wang2025spectralsimilaritymasksstructural}
\bibinfo{author}{Wang, Y.} \emph{et~al.}
\newblock \bibinfo{title}{Spectral similarity masks structural diversity at hydrophobic water interfaces} (\bibinfo{year}{2025}).
\newblock \bibinfo{eprint}{{\href{https://arxiv.org/abs/2504.05593}{{arXiv:2504.05593}}}}.

\bibitem{du2025machinelearningacceleratedcomputational}
\bibinfo{author}{Du, X.}, \bibinfo{author}{Cheng, J.} \& \bibinfo{author}{Tang, F.}
\newblock \bibinfo{title}{Machine learning accelerated computational surface-specific vibrational spectroscopy reveals oxidation level of graphene in contact with water} (\bibinfo{year}{2025}).
\newblock \bibinfo{eprint}{{\href{https://arxiv.org/abs/2507.00364}{{arXiv:2507.00364}}}}.

\bibitem{RN4}
\bibinfo{author}{Drautz, R.}
\newblock \bibinfo{title}{Atomic cluster expansion for accurate and transferable interatomic potentials}.
\newblock \emph{\bibinfo{journal}{Phys. Rev. B}} \textbf{\bibinfo{volume}{99}}, \bibinfo{pages}{014104} (\bibinfo{year}{2019}).

\bibitem{RN22}
\bibinfo{author}{Drautz, R.}
\newblock \bibinfo{title}{Atomic cluster expansion of scalar, vectorial, and tensorial properties including magnetism and charge transfer}.
\newblock \emph{\bibinfo{journal}{Phys. Rev. B}} \textbf{\bibinfo{volume}{102}}, \bibinfo{pages}{024104} (\bibinfo{year}{2020}).

\bibitem{PhysRevLett.122.225701}
\bibinfo{author}{Jinnouchi, R.}, \bibinfo{author}{Lahnsteiner, J.}, \bibinfo{author}{Karsai, F.}, \bibinfo{author}{Kresse, G.} \& \bibinfo{author}{Bokdam, M.}
\newblock \bibinfo{title}{Phase transitions of hybrid perovskites simulated by machine-learning force fields trained on the fly with bayesian inference}.
\newblock \emph{\bibinfo{journal}{Phys. Rev. Lett.}} \textbf{\bibinfo{volume}{122}}, \bibinfo{pages}{225701} (\bibinfo{year}{2019}).

\bibitem{PhysRevB.100.014105}
\bibinfo{author}{Jinnouchi, R.}, \bibinfo{author}{Karsai, F.} \& \bibinfo{author}{Kresse, G.}
\newblock \bibinfo{title}{On-the-fly machine learning force field generation: Application to melting points}.
\newblock \emph{\bibinfo{journal}{Phys. Rev. B}} \textbf{\bibinfo{volume}{100}}, \bibinfo{pages}{014105} (\bibinfo{year}{2019}).

\bibitem{10.1063/5.0009491}
\bibinfo{author}{Jinnouchi, R.}, \bibinfo{author}{Karsai, F.}, \bibinfo{author}{Verdi, C.}, \bibinfo{author}{Asahi, R.} \& \bibinfo{author}{Kresse, G.}
\newblock \bibinfo{title}{Descriptors representing two- and three-body atomic distributions and their effects on the accuracy of machine-learned inter-atomic potentials}.
\newblock \emph{\bibinfo{journal}{J. Chem. Phys.}} \textbf{\bibinfo{volume}{152}}, \bibinfo{pages}{234102} (\bibinfo{year}{2020}).

\bibitem{lysogorskiy2021performant}
\bibinfo{author}{Lysogorskiy, Y.} \emph{et~al.}
\newblock \bibinfo{title}{Performant implementation of the atomic cluster expansion (pace) and application to copper and silicon}.
\newblock \emph{\bibinfo{journal}{npj Comput. Mater.}} \textbf{\bibinfo{volume}{7}}, \bibinfo{pages}{97} (\bibinfo{year}{2021}).

\bibitem{PhysRevMaterials.7.043801}
\bibinfo{author}{Lysogorskiy, Y.}, \bibinfo{author}{Bochkarev, A.}, \bibinfo{author}{Mrovec, M.} \& \bibinfo{author}{Drautz, R.}
\newblock \bibinfo{title}{Active learning strategies for atomic cluster expansion models}.
\newblock \emph{\bibinfo{journal}{Phys. Rev. Mater.}} \textbf{\bibinfo{volume}{7}}, \bibinfo{pages}{043801} (\bibinfo{year}{2023}).

\bibitem{RN665}
\bibinfo{author}{Schultz, M.~J.}, \bibinfo{author}{Baldelli, S.}, \bibinfo{author}{Schnitzer, C.} \& \bibinfo{author}{Simonelli, D.}
\newblock \bibinfo{title}{Aqueous solution/air interfaces probed with sum frequency generation spectroscopy}.
\newblock \emph{\bibinfo{journal}{J. Phys. Chem. B}} \textbf{\bibinfo{volume}{106}}, \bibinfo{pages}{5313--5324} (\bibinfo{year}{2002}).

\bibitem{doi:10.1073/pnas.1906243117}
\bibinfo{author}{Niu, K.} \& \bibinfo{author}{Marcus, R.~A.}
\newblock \bibinfo{title}{Sum frequency generation, calculation of absolute intensities, comparison with experiments, and two-field relaxation-based derivation}.
\newblock \emph{\bibinfo{journal}{Proc. Natl. Acad. Sci. U.S.A.}} \textbf{\bibinfo{volume}{117}}, \bibinfo{pages}{2805--2814} (\bibinfo{year}{2020}).

\bibitem{10.1063/1.3135147}
\bibinfo{author}{Nihonyanagi, S.}, \bibinfo{author}{Yamaguchi, S.} \& \bibinfo{author}{Tahara, T.}
\newblock \bibinfo{title}{Direct evidence for orientational flip-flop of water molecules at charged interfaces: A heterodyne-detected vibrational sum frequency generation study}.
\newblock \emph{\bibinfo{journal}{J. Chem. Phys.}} \textbf{\bibinfo{volume}{130}}, \bibinfo{pages}{204704} (\bibinfo{year}{2009}).

\bibitem{RN666}
\bibinfo{author}{Tian, C.-S.} \& \bibinfo{author}{Shen, Y.~R.}
\newblock \bibinfo{title}{Isotopic dilution study of the water/vapor interface by phase-sensitive sum-frequency vibrational spectroscopy}.
\newblock \emph{\bibinfo{journal}{J. Am. Chem. Soc.}} \textbf{\bibinfo{volume}{131}}, \bibinfo{pages}{2790--2791} (\bibinfo{year}{2009}).

\bibitem{RN442}
\bibinfo{author}{Medders, G.~R.} \& \bibinfo{author}{Paesani, F.}
\newblock \bibinfo{title}{Dissecting the molecular structure of the air/water interface from quantum simulations of the sum-frequency generation spectrum}.
\newblock \emph{\bibinfo{journal}{J. Am. Chem. Soc.}} \textbf{\bibinfo{volume}{138}}, \bibinfo{pages}{3912--3919} (\bibinfo{year}{2016}).

\bibitem{annurev:/content/journals/10.1146/annurev-physchem-090722-124705}
\bibinfo{author}{Althorpe, S.~C.}
\newblock \bibinfo{title}{Path integral simulations of condensed-phase vibrational spectroscopy}.
\newblock \emph{\bibinfo{journal}{Annu. Rev. Phys. Chem.}} \textbf{\bibinfo{volume}{75}}, \bibinfo{pages}{397--420} (\bibinfo{year}{2024}).

\bibitem{RN75}
\bibinfo{author}{Koma, A.}, \bibinfo{author}{Saiki, K.} \& \bibinfo{author}{Sato, Y.}
\newblock \bibinfo{title}{Heteroepitaxy of a two-dimensional material on a three-dimensional material}.
\newblock \emph{\bibinfo{journal}{Appl. Surf. Sci.}} \textbf{\bibinfo{volume}{41-42}}, \bibinfo{pages}{451--456} (\bibinfo{year}{1990}).

\bibitem{RN650}
\bibinfo{author}{Mayerhöfer, T.~G.}, \bibinfo{author}{Pahlow, S.}, \bibinfo{author}{Hübner, U.} \& \bibinfo{author}{Popp, J.}
\newblock \bibinfo{title}{Caf2: An ideal substrate material for infrared spectroscopy?}
\newblock \emph{\bibinfo{journal}{Anal. Chem.}} \textbf{\bibinfo{volume}{92}}, \bibinfo{pages}{9024--9031} (\bibinfo{year}{2020}).

\bibitem{RN63}
\bibinfo{author}{Song, H.} \emph{et~al.}
\newblock \bibinfo{title}{Enhanced transport and optoelectronic properties of van der waals materials on caf2 films}.
\newblock \emph{\bibinfo{journal}{Nano Lett.}} \textbf{\bibinfo{volume}{23}}, \bibinfo{pages}{4983--4990} (\bibinfo{year}{2023}).

\bibitem{RN522}
\bibinfo{author}{Cen, J.} \emph{et~al.}
\newblock \bibinfo{title}{Adsorption of water molecule on calcium fluoride and magnesium fluoride surfaces: A combined theoretical and experimental study}.
\newblock \emph{\bibinfo{journal}{J. Phys. Chem. C}} \textbf{\bibinfo{volume}{124}}, \bibinfo{pages}{7853--7859} (\bibinfo{year}{2020}).

\bibitem{RN651}
\bibinfo{author}{de~Leeuw, N.~H.}, \bibinfo{author}{Purton, J.~A.}, \bibinfo{author}{Parker, S.~C.}, \bibinfo{author}{Watson, G.~W.} \& \bibinfo{author}{Kresse, G.}
\newblock \bibinfo{title}{Density functional theory calculations of adsorption of water at calcium oxide and calcium fluoride surfaces}.
\newblock \emph{\bibinfo{journal}{Surf. Sci.}} \textbf{\bibinfo{volume}{452}}, \bibinfo{pages}{9--19} (\bibinfo{year}{2000}).

\bibitem{RN652}
\bibinfo{author}{Yang, N.} \emph{et~al.}
\newblock \bibinfo{title}{Preparation of caf2 microspheres with nanopetals for water vapor adsorption}.
\newblock \emph{\bibinfo{journal}{Langmuir}} \textbf{\bibinfo{volume}{36}}, \bibinfo{pages}{5369--5376} (\bibinfo{year}{2020}).

\bibitem{C5SM02143J}
\bibinfo{author}{Chanda, J.}, \bibinfo{author}{Ionov, L.}, \bibinfo{author}{Kirillova, A.} \& \bibinfo{author}{Synytska, A.}
\newblock \bibinfo{title}{New insight into icing and de-icing properties of hydrophobic and hydrophilic structured surfaces based on core–shell particles}.
\newblock \emph{\bibinfo{journal}{Soft Matter}} \textbf{\bibinfo{volume}{11}}, \bibinfo{pages}{9126--9134} (\bibinfo{year}{2015}).

\bibitem{RN656}
\bibinfo{author}{Chen, L.}, \bibinfo{author}{He, X.}, \bibinfo{author}{Liu, H.}, \bibinfo{author}{Qian, L.} \& \bibinfo{author}{Kim, S.~H.}
\newblock \bibinfo{title}{Water adsorption on hydrophilic and hydrophobic surfaces of silicon}.
\newblock \emph{\bibinfo{journal}{J. Phys. Chem. C}} \textbf{\bibinfo{volume}{122}}, \bibinfo{pages}{11385--11391} (\bibinfo{year}{2018}).

\bibitem{10.1115/1.2826087}
\bibinfo{author}{Majumdar, A.} \& \bibinfo{author}{Mezic, I.}
\newblock \bibinfo{title}{Instability of ultra-thin water films and the mechanism of droplet formation on hydrophilic surfaces}.
\newblock \emph{\bibinfo{journal}{J. Heat Transfer}} \textbf{\bibinfo{volume}{121}}, \bibinfo{pages}{964--971} (\bibinfo{year}{1999}).

\bibitem{li2015two}
\bibinfo{author}{Li, Q.}, \bibinfo{author}{Song, J.}, \bibinfo{author}{Besenbacher, F.} \& \bibinfo{author}{Dong, M.}
\newblock \bibinfo{title}{Two-dimensional material confined water}.
\newblock \emph{\bibinfo{journal}{Acc. Chem. Res.}} \textbf{\bibinfo{volume}{48}}, \bibinfo{pages}{119--127} (\bibinfo{year}{2015}).

\bibitem{dollekamp2017charge}
\bibinfo{author}{Dollekamp, E.}, \bibinfo{author}{Bampoulis, P.}, \bibinfo{author}{Faasen, D.~P.}, \bibinfo{author}{Zandvliet, H.~J.} \& \bibinfo{author}{Kooij, E.~S.}
\newblock \bibinfo{title}{Charge induced dynamics of water in a graphene--mica slit pore}.
\newblock \emph{\bibinfo{journal}{Langmuir}} \textbf{\bibinfo{volume}{33}}, \bibinfo{pages}{11977--11985} (\bibinfo{year}{2017}).

\bibitem{okmi2022discovery}
\bibinfo{author}{Okmi, A.} \emph{et~al.}
\newblock \bibinfo{title}{Discovery of graphene-water membrane structure: Toward high-quality graphene process}.
\newblock \emph{\bibinfo{journal}{Adv. Sci.}} \textbf{\bibinfo{volume}{9}}, \bibinfo{pages}{2201336} (\bibinfo{year}{2022}).

\bibitem{RN526}
\bibinfo{author}{Verdaguer, A.}, \bibinfo{author}{Segura, J.~J.}, \bibinfo{author}{López-Mir, L.}, \bibinfo{author}{Sauthier, G.} \& \bibinfo{author}{Fraxedas, J.}
\newblock \bibinfo{title}{Communication: Growing room temperature ice with graphene}.
\newblock \emph{\bibinfo{journal}{J. Chem. Phys.}} \textbf{\bibinfo{volume}{138}}, \bibinfo{pages}{121101} (\bibinfo{year}{2013}).

\bibitem{stiopkin2011hydrogen}
\bibinfo{author}{Stiopkin, I.~V.} \emph{et~al.}
\newblock \bibinfo{title}{Hydrogen bonding at the water surface revealed by isotopic dilution spectroscopy}.
\newblock \emph{\bibinfo{journal}{Nature}} \textbf{\bibinfo{volume}{474}}, \bibinfo{pages}{192--195} (\bibinfo{year}{2011}).

\bibitem{RN649}
\bibinfo{author}{Yavari, F.} \emph{et~al.}
\newblock \bibinfo{title}{Tunable bandgap in graphene by the controlled adsorption of water molecules}.
\newblock \emph{\bibinfo{journal}{Small}} \textbf{\bibinfo{volume}{6}}, \bibinfo{pages}{2535--2538} (\bibinfo{year}{2010}).

\bibitem{RN664}
\bibinfo{author}{Chen, J.~H.} \emph{et~al.}
\newblock \bibinfo{title}{Charged-impurity scattering in graphene}.
\newblock \emph{\bibinfo{journal}{Nat. Phys.}} \textbf{\bibinfo{volume}{4}}, \bibinfo{pages}{377--381} (\bibinfo{year}{2008}).

\bibitem{wang2025spontaneoussurfacechargingjanus}
\bibinfo{author}{Wang, Y.} \emph{et~al.}
\newblock \bibinfo{title}{Spontaneous surface charging and janus nature of the hexagonal boron nitride–water interface}.
\newblock \emph{\bibinfo{journal}{J. Am. Chem. Soc.}} \textbf{\bibinfo{volume}{147}}, \bibinfo{pages}{30107--30116} (\bibinfo{year}{2025}).

\bibitem{10.1063/1.3033202}
\bibinfo{author}{Wehling, T.~O.}, \bibinfo{author}{Lichtenstein, A.~I.} \& \bibinfo{author}{Katsnelson, M.~I.}
\newblock \bibinfo{title}{First-principles studies of water adsorption on graphene: The role of the substrate}.
\newblock \emph{\bibinfo{journal}{Appl. Phys. Lett.}} \textbf{\bibinfo{volume}{93}}, \bibinfo{pages}{202110} (\bibinfo{year}{2008}).

\bibitem{PhysRevB.47.558}
\bibinfo{author}{Kresse, G.} \& \bibinfo{author}{Hafner, J.}
\newblock \bibinfo{title}{Ab initio molecular dynamics for liquid metals}.
\newblock \emph{\bibinfo{journal}{Phys. Rev. B}} \textbf{\bibinfo{volume}{47}}, \bibinfo{pages}{558--561} (\bibinfo{year}{1993}).

\bibitem{KRESSE199615}
\bibinfo{author}{Kresse, G.} \& \bibinfo{author}{Furthmüller, J.}
\newblock \bibinfo{title}{Efficiency of ab-initio total energy calculations for metals and semiconductors using a plane-wave basis set}.
\newblock \emph{\bibinfo{journal}{Comput. Mater. Sci.}} \textbf{\bibinfo{volume}{6}}, \bibinfo{pages}{15--50} (\bibinfo{year}{1996}).

\bibitem{PhysRevB.54.11169}
\bibinfo{author}{Kresse, G.} \& \bibinfo{author}{Furthm\"uller, J.}
\newblock \bibinfo{title}{Efficient iterative schemes for ab initio total-energy calculations using a plane-wave basis set}.
\newblock \emph{\bibinfo{journal}{Phys. Rev. B}} \textbf{\bibinfo{volume}{54}}, \bibinfo{pages}{11169--11186} (\bibinfo{year}{1996}).

\bibitem{PhysRevB.59.1758}
\bibinfo{author}{Kresse, G.} \& \bibinfo{author}{Joubert, D.}
\newblock \bibinfo{title}{From ultrasoft pseudopotentials to the projector augmented-wave method}.
\newblock \emph{\bibinfo{journal}{Phys. Rev. B}} \textbf{\bibinfo{volume}{59}}, \bibinfo{pages}{1758--1775} (\bibinfo{year}{1999}).

\bibitem{PhysRevB.50.17953}
\bibinfo{author}{Bl\"ochl, P.~E.}
\newblock \bibinfo{title}{Projector augmented-wave method}.
\newblock \emph{\bibinfo{journal}{Phys. Rev. B}} \textbf{\bibinfo{volume}{50}}, \bibinfo{pages}{17953--17979} (\bibinfo{year}{1994}).

\bibitem{PhysRevA.38.3098}
\bibinfo{author}{Becke, A.~D.}
\newblock \bibinfo{title}{Density-functional exchange-energy approximation with correct asymptotic behavior}.
\newblock \emph{\bibinfo{journal}{Phys. Rev. A}} \textbf{\bibinfo{volume}{38}}, \bibinfo{pages}{3098--3100} (\bibinfo{year}{1988}).

\bibitem{PhysRevB.37.785}
\bibinfo{author}{Lee, C.}, \bibinfo{author}{Yang, W.} \& \bibinfo{author}{Parr, R.~G.}
\newblock \bibinfo{title}{Development of the colle-salvetti correlation-energy formula into a functional of the electron density}.
\newblock \emph{\bibinfo{journal}{Phys. Rev. B}} \textbf{\bibinfo{volume}{37}}, \bibinfo{pages}{785--789} (\bibinfo{year}{1988}).

\bibitem{10.1063/1.5090222}
\bibinfo{author}{Caldeweyher, E.} \emph{et~al.}
\newblock \bibinfo{title}{A generally applicable atomic-charge dependent london dispersion correction}.
\newblock \emph{\bibinfo{journal}{J. Chem. Phys.}} \textbf{\bibinfo{volume}{150}}, \bibinfo{pages}{154122} (\bibinfo{year}{2019}).

\bibitem{PhysRevB.17.1302}
\bibinfo{author}{Schneider, T.} \& \bibinfo{author}{Stoll, E.}
\newblock \bibinfo{title}{Molecular-dynamics study of a three-dimensional one-component model for distortive phase transitions}.
\newblock \emph{\bibinfo{journal}{Phys. Rev. B}} \textbf{\bibinfo{volume}{17}}, \bibinfo{pages}{1302--1322} (\bibinfo{year}{1978}).

\bibitem{bochkarev2022efficient}
\bibinfo{author}{Bochkarev, A.} \emph{et~al.}
\newblock \bibinfo{title}{Efficient parametrization of the atomic cluster expansion}.
\newblock \emph{\bibinfo{journal}{Phys. Rev. Mater.}} \textbf{\bibinfo{volume}{6}}, \bibinfo{pages}{013804} (\bibinfo{year}{2022}).

\bibitem{10.1063/1.2408420}
\bibinfo{author}{Bussi, G.}, \bibinfo{author}{Donadio, D.} \& \bibinfo{author}{Parrinello, M.}
\newblock \bibinfo{title}{Canonical sampling through velocity rescaling}.
\newblock \emph{\bibinfo{journal}{J. Chem. Phys.}} \textbf{\bibinfo{volume}{126}}, \bibinfo{pages}{014101} (\bibinfo{year}{2007}).

\bibitem{doi:10.1142/9789812836021_0015}
\bibinfo{author}{Goreinov, S.~A.}, \bibinfo{author}{Oseledets, I.~V.}, \bibinfo{author}{Savostyanov, D.~V.}, \bibinfo{author}{Tyrtyshnikov, E.~E.} \& \bibinfo{author}{Zamarashkin, N.~L.}
\newblock \emph{\bibinfo{title}{How to Find a Good Submatrix}}, \bibinfo{pages}{247--256} (\bibinfo{publisher}{World Scientific}, \bibinfo{address}{Singapore}, \bibinfo{year}{2010}).

\bibitem{LAMMPS}
\bibinfo{author}{Thompson, A.~P.} \emph{et~al.}
\newblock \bibinfo{title}{{LAMMPS} - a flexible simulation tool for particle-based materials modeling at the atomic, meso, and continuum scales}.
\newblock \emph{\bibinfo{journal}{Comp. Phys. Comm.}} \textbf{\bibinfo{volume}{271}}, \bibinfo{pages}{108171} (\bibinfo{year}{2022}).

\bibitem{PLIMPTON19951}
\bibinfo{author}{Plimpton, S.}
\newblock \bibinfo{title}{Fast parallel algorithms for short-range molecular dynamics}.
\newblock \emph{\bibinfo{journal}{J. Comput. Phys.}} \textbf{\bibinfo{volume}{117}}, \bibinfo{pages}{1--19} (\bibinfo{year}{1995}).

\bibitem{GOREINOV19971}
\bibinfo{author}{Goreinov, S.}, \bibinfo{author}{Tyrtyshnikov, E.} \& \bibinfo{author}{Zamarashkin, N.}
\newblock \bibinfo{title}{A theory of pseudoskeleton approximations}.
\newblock \emph{\bibinfo{journal}{Linear Algebra Its Appl.}} \textbf{\bibinfo{volume}{261}}, \bibinfo{pages}{1--21} (\bibinfo{year}{1997}).

\bibitem{10.1063/1.5016629}
\bibinfo{author}{Yesudas, F.}, \bibinfo{author}{Mero, M.}, \bibinfo{author}{Kneipp, J.} \& \bibinfo{author}{Heiner, Z.}
\newblock \bibinfo{title}{Vibrational sum-frequency generation spectroscopy of lipid bilayers at repetition rates up to 100 khz}.
\newblock \emph{\bibinfo{journal}{J. Chem. Phys.}} \textbf{\bibinfo{volume}{148}}, \bibinfo{pages}{104702} (\bibinfo{year}{2018}).

\bibitem{KIM20221187}
\bibinfo{author}{Kim, E.} \emph{et~al.}
\newblock \bibinfo{title}{Wettability of graphene, water contact angle, and interfacial water structure}.
\newblock \emph{\bibinfo{journal}{Chem}} \textbf{\bibinfo{volume}{8}}, \bibinfo{pages}{1187--1200} (\bibinfo{year}{2022}).

\bibitem{D0CC02675A}
\bibinfo{author}{AlSalem, H.~S.} \emph{et~al.}
\newblock \bibinfo{title}{Imaging the reactivity and width of graphene{'}s boundary region}.
\newblock \emph{\bibinfo{journal}{Chem. Commun.}} \textbf{\bibinfo{volume}{56}}, \bibinfo{pages}{9612--9615} (\bibinfo{year}{2020}).

\bibitem{RN559}
\bibinfo{author}{Medders, G.~R.} \& \bibinfo{author}{Paesani, F.}
\newblock \bibinfo{title}{Infrared and raman spectroscopy of liquid water through “first-principles” many-body molecular dynamics}.
\newblock \emph{\bibinfo{journal}{J. Chem. Theory Comput.}} \textbf{\bibinfo{volume}{11}}, \bibinfo{pages}{1145--1154} (\bibinfo{year}{2015}).

\bibitem{10.1063/1.4916629}
\bibinfo{author}{Medders, G.~R.} \& \bibinfo{author}{Paesani, F.}
\newblock \bibinfo{title}{On the interplay of the potential energy and dipole moment surfaces in controlling the infrared activity of liquid water}.
\newblock \emph{\bibinfo{journal}{J. Chem. Phys.}} \textbf{\bibinfo{volume}{142}}, \bibinfo{pages}{212411} (\bibinfo{year}{2015}).

\bibitem{10.1063/1.5006480}
\bibinfo{author}{Reddy, S.~K.}, \bibinfo{author}{Moberg, D.~R.}, \bibinfo{author}{Straight, S.~C.} \& \bibinfo{author}{Paesani, F.}
\newblock \bibinfo{title}{Temperature-dependent vibrational spectra and structure of liquid water from classical and quantum simulations with the mb-pol potential energy function}.
\newblock \emph{\bibinfo{journal}{J. Chem. Phys.}} \textbf{\bibinfo{volume}{147}}, \bibinfo{pages}{244504} (\bibinfo{year}{2017}).

\bibitem{RN233}
\bibinfo{author}{Moberg, D.~R.}, \bibinfo{author}{Straight, S.~C.} \& \bibinfo{author}{Paesani, F.}
\newblock \bibinfo{title}{Temperature dependence of the air/water interface revealed by polarization sensitive sum-frequency generation spectroscopy}.
\newblock \emph{\bibinfo{journal}{J. Phys. Chem. B}} \textbf{\bibinfo{volume}{122}}, \bibinfo{pages}{4356} (\bibinfo{year}{2018}).

\bibitem{RN167}
\bibinfo{author}{Nagata, Y.} \& \bibinfo{author}{Mukamel, S.}
\newblock \bibinfo{title}{Vibrational sum-frequency generation spectroscopy at the water/lipid interface: Molecular dynamics simulation study}.
\newblock \emph{\bibinfo{journal}{J. Am. Chem. Soc.}} \textbf{\bibinfo{volume}{132}}, \bibinfo{pages}{6434} (\bibinfo{year}{2010}).

\bibitem{RN113}
\bibinfo{author}{Tang, F.} \emph{et~al.}
\newblock \bibinfo{title}{Molecular structure and modeling of water–air and ice–air interfaces monitored by sum-frequency generation}.
\newblock \emph{\bibinfo{journal}{Chem. Rev.}} \textbf{\bibinfo{volume}{120}}, \bibinfo{pages}{3633--3667} (\bibinfo{year}{2020}).

\bibitem{10.1116/6.0001401}
\bibinfo{author}{Pickering, J.~D.} \emph{et~al.}
\newblock \bibinfo{title}{Tutorials in vibrational sum frequency generation spectroscopy. i. the foundations}.
\newblock \emph{\bibinfo{journal}{Biointerphases}} \textbf{\bibinfo{volume}{17}}, \bibinfo{pages}{011201} (\bibinfo{year}{2022}).

\bibitem{RN569}
\bibinfo{author}{Lin, S.-T.}, \bibinfo{author}{Blanco, M.} \& \bibinfo{author}{Goddard, I., William~A.}
\newblock \bibinfo{title}{The two-phase model for calculating thermodynamic properties of liquids from molecular dynamics: Validation for the phase diagram of lennard-jones fluids}.
\newblock \emph{\bibinfo{journal}{J. Chem. Phys.}} \textbf{\bibinfo{volume}{119}}, \bibinfo{pages}{11792--11805} (\bibinfo{year}{2003}).

\bibitem{RN570}
\bibinfo{author}{Caro, M.~A.}, \bibinfo{author}{Laurila, T.} \& \bibinfo{author}{Lopez-Acevedo, O.}
\newblock \bibinfo{title}{Accurate schemes for calculation of thermodynamic properties of liquid mixtures from molecular dynamics simulations}.
\newblock \emph{\bibinfo{journal}{J. Chem. Phys.}} \textbf{\bibinfo{volume}{145}}, \bibinfo{pages}{244504} (\bibinfo{year}{2016}).

\end{thebibliography}

\end{document}